\documentclass[british]{zarticle}
\usepackage{CJK}
\usepackage{tikz}
\usepackage{arydshln}
\usetikzlibrary{decorations.pathreplacing}

\newcommand{\rd}{\mathrm{d}}
\newcommand{\rT}{\mathrm{T}}
\newcommand{\br}{\mathbf{r}}
\newcommand{\cH}{\mathcal{H}}
\newcommand{\cP}{\mathcal{P}}
\newcommand{\cR}{\mathcal{R}}
\newcommand{\cZ}{\mathcal{Z}}
\newcommand{\BI}{\mathbb{I}}
\newcommand{\BR}{\mathbb{R}}
\newcommand{\BZ}{\mathbb{Z}}

\newcommand{\SO}{\mathrm{SO}}
\newcommand{\Sp}{\mathrm{Sp}}
\newcommand{\SU}{\mathrm{SU}}
\newcommand{\rO}{\mathrm{O}}
\newcommand{\rU}{\mathrm{U}}

\title{Advancing Fuzzy-Sphere CFTs:} 
\subtitle{$\SO(3)$-Rotation-Resolving Exact Diagonalization, 3D Ising Heavy Primaries, and Quasi-Hole-Space Projection}
\date{29th September 2026}
\author{Zheng Zhou (周正)}
\author{Yin-Chen He}
\affiliation{Perimeter Institute for Theoretical Physics, Waterloo, Ontario N2L 2Y5, Canada}

\abstract{Fuzzy-sphere regularization has recently emerged as a powerful framework for studying three-dimensional conformal field theories (CFTs). In this work, we exploit symmetry and Hilbert-space structures to substantially advance this framework. We develop an exact-diagonalization algorithm that resolves the full $\SO(3)$ rotational symmetry, reducing computational costs by factors of several hundred. For the three-dimensional Ising CFT, we identify around 100 primary operators, including a large number of previously unknown heavy scalar primaries up to scaling dimension $\Delta\approx16.5$. We also resolve the tension between previous fuzzy-sphere and bootstrap results regarding the parity-odd scalar primary. Furthermore, we identify a branch of scalar primaries associated with $\phi^n$ operators in the $\phi^4$ Lagrangian description, beginning with the identity, $\sigma$, $\epsilon$, $\epsilon'$, and $\sigma'$. Their wave functions exhibit a semi-classical structure closely related to states invariant under area-preserving diffeomorphisms of the sphere. Finally, we show that critical points involving fractional quantum Hall states can be studied within a substantially reduced Hilbert space spanned by quasi-hole states of the corresponding model quantum Hall wave functions like the Laughlin state. This suggests a general strategy for exploring CFTs intertwined with topologically ordered phases.}

\begin{document}

\begin{CJK*}{UTF8}{bkai}
\maketitle
\end{CJK*}

\section{Introduction}

Conformal field theory (CFT) has provided important insights into condensed-matter~\cite{Polyakov1970Conformal,Cardy1996Scaling,Sachdev2011Quantum} and high-energy~\cite{Polchinski1998String,Maldacena1998AdSCFT,Zamolodchikov1986Irreversibility} physics. Compared with 2D CFTs~\cite{DiFrancesco1997CFT,Ginsparg1988CFT,Belavin1984BPZ}, the CFTs in higher space-time dimensions are less well-studied due to a much smaller conformal group. A paradigmatic example is the 3D Ising CFT that captures the paramagnet-ferromagnet critical point in the 3D Ising model. As one of the simplest instances of interacting CFTs, it has served as a test bed for methods developed to study 3D CFTs, including conformal bootstrap~\cite{Poland2018Bootstrap,Rychkov2023Bootstrap}, Monte Carlo~\cite{Hasenbusch:2017yxr,Ferrenberg2018IsingMC,Hasenbusch:2021tei}, and perturbative methods like $\epsilon$-expansion~\cite{Wilson:1971dc,Henriksson:2025vyi,Shapoval:2026iwc}, which have achieved considerable success in extracting conformal data, including scaling dimensions and operator-product-expansion (OPE) coefficients, for the leading operators. For a review of the methods and conformal data of $\rO(N)$ CFTs in dimensions $2<d<4$, including the Ising case $N=1$, see Ref.~\cite{Henriksson:2022rnm}.

The fuzzy-sphere regularization~\cite{Zhu2022,He2026} has emerged as a new powerful method to study 3D CFTs. This method involves studying quantum systems on a sphere rendered ``fuzzy''~\cite{Madore1991Fuzzy}, \emph{i.~e.}~non-commutative, by projection onto the lowest Landau level in the magnetic field generated by a monopole at its centre~\cite{Haldane1983FQHE,Ippoliti:2018ojo}. It offers distinct advantages including the exact preservation of rotational symmetry, the direct observation of emergent conformal symmetry and the efficient extraction of conformal data. In the fuzzy-sphere method, the state-operator correspondence plays an essential role. Specifically, there is a one-to-one correspondence between the eigenstates of the critical Hamiltonian on the sphere and the CFT operators, where the energy gaps are proportional to the scaling dimensions. The power of this approach has been first demonstrated in the context of the 3D Ising transition~\cite{Zhu2022}, where the presence of emergent conformal symmetry has been convincingly established. 

The fuzzy-sphere method has provided access to various properties of 3D CFTs since its proposal, such as the OPE coefficients~\cite{Hu2023Mar,Fardelli2026}, correlation functions~\cite{Han2023Jun}, entropic $F$-function~\cite{Hu2024}, conformal generators~\cite{Fardelli2024,Fan2024,Fardelli2026}, conformal perturbation~\cite{Laeuchli2025}, \textit{etc.}~\cite{Voinea2024,Wiese2025,Eck2026,Janssens2026,Zeng2026,Dong2025} for the Ising CFT. It has also been extended to a wide range of 3D CFTs~\cite{Zhou2023,Han2023Dec,Zhou2024Oct,Yang2025Jan,Fan2025,ArguelloCruz2025,EliasMiro2025,He2025Jun,Taylor2025,Yang2025Jul,Zhou2025Jul,Zhou2025Sep,Voinea2025,Dey2025,Guo2025,Tang2025,Huffman2026,Dey2026,Stergiou2026} and their defects~\cite{Hu2023Aug,Zhou2024Jan,Cuomo2024,Zhou2024Jul,Dedushenko2024,Sarma2026,Feng2026}, as well as CFTs in other dimensions~\cite{Han2025,Zhao2025,Meng2026,Xue2026,Qian2026Sep}. It has opened up a broad new territory for exploring 3D CFTs. Many of their properties are difficult to access with other methods such as the conformal bootstrap and Monte Carlo simulations.

One particularly compelling frontier is the heavy operators of 3D CFTs. Heavy operators probe a qualitatively different regime from the low-lying spectrum that has been the focus of most quantitative studies, and are closely connected to broader questions of quantum chaos, thermalization, holography, and the physics of black-hole microstates~\cite{Maldacena:2015waa,Lashkari:2016vgj,Fitzpatrick:2015zha}. Despite this motivation, the data related to heavy operators remain challenging to access with existing methods such as Monte Carlo simulations and conformal bootstrap.

The fuzzy-sphere approach provides a natural route to this regime: Through the state-operator correspondence, heavy operators are mapped directly to high-lying excited states on the sphere. Access to these states has so far been limited by the available numerical algorithms. Density matrix renormalization group (DMRG)~\cite{Hu2023Mar,Hao2026} and quantum Monte Carlo (QMC)~\cite{Hofmann2023,Huffman2026} can reach relatively large system sizes, but provide access only to a few low-lying excited states. Exact diagonalization (ED), such as the Lanczos algorithm, can resolve a substantially larger number of excited states. Existing fuzzy-sphere ED algorithms resolve only the $\rU(1)$ subgroup of the full $\SO(3)$ rotational symmetry of the sphere, limiting their reach to higher-lying states. Because a fixed $\rU(1)$ sector contains states from many different $\SO(3)$ representations, much of the computation effort is spent resolving states outside the target angular-momentum sector. On the other hand, methods for constructing $\SO(3)$-resolved Hilbert spaces through recursive coupling of electrons or orbitals have long been used in atomic physics and quantum chemistry~\cite{Racah1943,Gaigalas2005}. As the number of electrons increases, constructing these bases and evaluating Hamiltonian matrix elements within them become computationally demanding.

In this paper, we introduce a new method resolving the full $\SO(3)$ rotational symmetry by partitioning the model into segments, resolving the $\SO(3)$ symmetry in each segment by fully diagonalizing the quadratic Casimir of the angular momentum $L^2$, and coupling the resulting segment spaces to form a basis with definite total angular momentum. This algorithm reduces memory costs by roughly two orders of magnitude, enabling substantially larger system sizes while dramatically extending the accessible excited state spectrum. Taking the 3D Ising CFT as an example, the existing $\rU(1)$-resolving algorithm gives access to only around 20 $\SO(3)$ scalars at its largest size, reaching scaling dimensions of roughly $\Delta=8$~\cite{Zhu2022,Fardelli2026}. In contrast, our $\SO(3)$-resolving ED algorithm allows us to access 4600 $\SO(3)$ scalars and extends the spectrum to scaling dimensions of roughly $\Delta=22$ (parity-even) or $19$ (parity-odd). Combining this algorithm with the conformal generators~\cite{Fardelli2024} --- in particular, the $|K|^2$ filtering technique~\cite{Fardelli2026} --- we identify a large number of previously unknown heavy primary operators in the 3D Ising CFT, reaching $\Delta\approx16.5$, as summarized in Table~\ref{tbl:prim_dim}. Notably, we find that the lowest parity-odd scalar primary was misidentified in previous fuzzy-sphere calculations~\cite{Zhu2022}, thereby resolving the tension with the stress-tensor bootstrap~\cite{Chang:2024whx}.

\newcommand{\CB}{^{[\text{CB}]}}
\newcommand{\FS}{^{[\text{FS}]}}
\newcommand{\EE}[1]{^{[\epsilon^#1]}}

\begin{table}[htbp]
    \renewcommand{\tabcolsep}{3pt}
    \newcommand{\er}[1]{_{(#1)}}
    \centering\small
    \begin{tabular}{cccc|llllllll}
        \hline\hline
        $l$ & $Z$ & $P$ & Method & \multicolumn{8}{c}{Scaling dimensions}\\\hline
        0 & $+$ & $+$ & Fuzzy & $1.416\er{8}$ & $3.86\er{3}$ & $6.96\er{5}$ & $7.1\er{2}$ & $8.3\er{3}$ & $10.43\er{9}$ & $10.5\er{2}$ & $11.2\er{5}$ \\
        & & & \emph{Other} & $\mathit{1.412}\CB$ & $\mathit{3.830}\CB$ & $\mathit{6.896}\CB$ & $\mathit{7.254}\CB$ & $\mathit{8.92}\CB$ & $\mathit{10.56}\EE5$ & $\mathit{10.19}\EE2$ & $\mathit{10.93}\EE1$ \\ \cdashline{5-12}[1pt/4pt]
        & & & Fuzzy & $11.5\er{7}$ & $11.7\er{4}$ & $12.2\er{7}$ & $12.6\er{7}$ & $13.4\er{8}$ & $14.0\er{3}$ & $14.0\er{9}$ & $14.1\er{6}$ \\
        & & & \emph{Other} & $\mathit{11.67}\EE1$ & $\mathit{11.67}\EE1$ & & & $\mathit{13.06}\EE2$ \\ \cdashline{5-12}[1pt/4pt]
        & & & Fuzzy & $14.2\er{10}$ & $14.68\er{6}$ & $14.8\er{1.2}$ & $15.0\er{6}$ & $15.5\er{6}$ \\
        & & & \emph{Other} & & $\mathit{14.98}\EE5$ & & & \\ \hline
        0 & $-$ & $+$ & Fuzzy & $0.518$ & $5.34\er{4}$ & $8.68\er{6}$ & $8.7\er{2}$ & $10.0\er{3}$ & $12.3\er{3}$ & $12.53\er{15}$ & $12.7\er{6}$ \\
        & & & \emph{Other} & $\mathit{0.518}\CB$ & $\mathit{5.291}\CB$ & $\mathit{8.627}\EE5$ & $\mathit{8.67}\EE3$ & $\mathit{9.83}\EE1$ & $\mathit{11.61}\EE2$ & $\mathit{12.67}\EE5$ & \\ \cdashline{5-12}[1pt/4pt]
        & & & Fuzzy & $13.3\er{6}$ & $13.5\er{6}$ & $13.7\er{1.1}$ & $14.3\er{7}$ & $14.4\er{1.3}$ & $14.9\er{1.4}$ & $15.3\er{10}$ & $16\er{2}$ \\ \cdashline{5-12}[1pt/4pt]
        & & & Fuzzy & $16.1\er{1.3}$ & $16.3\er{7}$ \\ \hline
        0 & $+$ & $-$ & Fuzzy & $11.0\er{4}$ & $13.3\er{6}$ & $14.8\er{7}$ & $14.8\er{1.8}$ & $15.9\er{1.1}$ & $16.2\er{1.8}$ \\
        & & & \emph{Other} & $\mathit{10.90}\er{3}\CB$ \\ \hline
        0 & $-$ & $-$ & Fuzzy & $12.8\er{7}$ & $15.1\er{8}$ & $16.49\er{10}$ \\ \hline
        1 & $+$ & $+$ & Fuzzy & $10.8\er{5}$ & $12.0\er{7}$ & $12.8\er{4}$ & $13.2\er{8}$ \\
        & & & \textit{Other} & $\mathit{11.11}\EE1$ \\ \hline
        1 & $-$ & $+$ & Fuzzy & $9.1\er{4}$ & $10.2\er{5}$ & $11.2\er{5}$ & $12.1\er{5}$ & $12.7\er{4}$ & $12.9\er{6}$ & $13.5\er{7}$ \\
        & & & \emph{Other} & $\mathit{9.4}\CB$ & & & & & \\ \hline
        1 & $+$ & $-$ & Fuzzy & $12.9\er{7}$ & $13.2\er{1.0}$ & $14.2\er{5}$ & $15.0\er{4}$ \\ \hline
        1 & $-$ & $-$ & Fuzzy & $11.1\er{6}$ & $12.9\er{5}$ & $13.6\er{5}$ & $15.0\er{4}$ \\
        \hline
        2 & $+$ & $+$ & Fuzzy & $2.991\er{14}$ & $5.549\er{13}$ & $7.03\er{17}$ & $8.70\er{6}$ & $9.1\er{4}$ & $10.29\er{15}$ & $10.3\er{4}$ & $11.1\er{4}$ \\
        & & & \emph{Other} & $\mathit{3}$ & $\mathit{5.509}\CB$ & $\mathit{7.076}\CB$  & $\mathit{8.35}\EE2$ & $\mathit{8.93}\EE1$ & $\mathit{10.44}\EE1$ & & $\mathit{11.21}\EE2$\\ \cdashline{5-12}[1pt/4pt]
        & & & Fuzzy & $11.2\er{4}$ & $11.3\er{4}$ & $12.29\er{9}$ \\ \hline
        2 & $-$ & $+$ & Fuzzy & $4.197\er{10}$ & $7.058\er{15}$ & $8.61\er{17}$ & $9.4\er{4}$ & $10.44\er{15}$ & $10.6\er{3}$ & $11.0\er{5}$ & $11.6\er{5}$ \\
        & & & \emph{Other} & $\mathit{4.180}\CB$ & $\mathit{6.987}\CB$ & $\mathit{8.50}\EE1$ & $\mathit{9.77}\EE2$ & & \\ \cdashline{5-12}[1pt/4pt]
        & & & Fuzzy & $12.07\er{16}$ & & \\ \hline
        2 & $+$ & $-$ & Fuzzy & $7.4\er{2}$ & $9.2\er{3}$ & $10.8\er{3}$ & $11.5\er{8}$ & $11.7\er{6}$ \\ 
        & & & \textit{Other} & $\mathit{7.50}\EE1$ \\ \hline
        2 & $-$ & $-$ & Fuzzy & $9.1\er{2}$ & $10.9\er{5}$ & $11.0\er{5}$ \\
        & & & \textit{Other} & $\mathit{9.06}\EE1$ \\ 
        \hline\hline
    \end{tabular}
    \caption{The scaling dimensions of the primary operators identified using the conformal generator. We compare our results with those obtained with other methods, \emph{viz.}~conformal bootstrap~\cite{Poland2018Bootstrap,Rychkov2023Bootstrap,Simmons-Duffin:2016wlq,Chang:2024whx,Henriksson:2022gpa} and $\epsilon$-expansion~\cite{Antipin:2025ilv,Henriksson:2025vyi}. The $\epsilon$-expansion results are marked by their order.}
    \label{tbl:prim_dim}
\end{table}

The $\SO(3)$-resolving algorithm also facilitates the investigation of principles organizing the CFT states within the fuzzy-sphere Hilbert space. Understanding these principles may, on the one hand, improve the efficiency of extracting previously unknown CFT data using the fuzzy-sphere method and, on the other hand, potentially point toward new theoretical structures underlying 3D CFTs.

For the Ising CFT, we identify a distinguished branch of scalar primaries associated with the $\phi^n$ family in the $\phi^4$ Lagrangian description, whose lowest members include the identity, $\sigma$, $\epsilon$, $\epsilon'$, and $\sigma'$. These states have unusually large wave-function weights in a small subspace invariant under area-preserving diffeomorphisms (APD) of the sphere. This subspace is purely classical, spanned by states obtained through collective spin flips from a fully polarized configuration.

Another discovery concerns critical points involving fractional quantum Hall (FQH) phases, such as the Ising CFT at fractional filling~\cite{Voinea2024} and the $\SU(2)_1$-Higgs transition~\cite{Zhou2025Jul}. These transitions are particularly interesting because of their connections to field-theory dualities~\cite{Senthil:2018cru} and experiments in Moir\'e materials~\cite{Cai2023Moire,Zeng2023Moire,Park2023Moire}. We find that the CFT states at these transitions lie predominantly within the subspaces spanned by the quasi-hole states of the corresponding FQH wave functions~\cite{Haldane1983FQHE}, such as the Laughlin~\cite{Laughlin1983Anomalous}, Halperin~\cite{Halperin:1983zz}, and Moore-Read~\cite{Moore1991Pfaffian} states. These quasi-hole spaces are often also the zero-energy space of the parent Hamiltonian of the FQH wave function. Projecting onto these quasi-hole subspaces substantially reduces the Hilbert-space dimension and, more importantly, yields a much cleaner CFT spectrum. This is particularly important because magneto-roton modes and other gapped excitations of the underlying FQH states can occur at relatively low energies and overlap with the CFT spectrum. These additional states are distinct from the finite-size shifts of CFT levels described by conformal perturbation theory. The quasi-hole-space projection removes most of these non-CFT states while improving the CFT spectrum.

The rest of this paper is organized as follows: in \S~\ref{sec:method}, we give a detailed prescription of the algorithm and benchmark its performance with previous $\rU(1)$-resolving ED; in \S~\ref{sec:results_prim}, we present the newly identified primaries; in \S~\ref{sec:results_apd}, we discuss the $\phi^n$-family of scalar primaries and their relation with APD; in \S~\ref{sec:results_trunc}, we discuss the quasi-hole-space projection of fractional quantum Hall transitions; finally, we summarize and discuss potential implications and prospective directions in \S~\ref{sec:discussion}.

\section{$\SO(3)$-Rotation-Resolving Exact Diagonalization}
\label{sec:method}

This section discusses the numerical algorithm for implementing the full $\SO(3)$ rotational symmetry on the sphere in the ED computation after a review of the fuzzy-sphere Ising model in \S~\ref{sec:model}. We first build the $\SO(3)$-resolved Hilbert space in \S~\ref{sec:method_hilbert}, and then build the Hamiltonian operator on the Hilbert space in \S\S~\ref{sec:method_operator} and \ref{sec:method_decomp}. After presenting the numerical procedure in \S~\ref{sec:method_summary}, we discuss the implementation of particle-hole symmetry in \S~\ref{sec:method_ph}, the generalization to other models in \S~\ref{sec:method_other}, benchmark the performance, and compare it with previous $\rU(1)$-resolving ED in \S~\ref{sec:method_benchmark}. Readers primarily interested in the results may wish to skip this section on a first reading.

\subsection{Fuzzy-Sphere Ising Model}
\label{sec:model}

As a representative example to demonstrate the algorithm, we consider the fuzzy-sphere model for the 3D Ising CFT~\cite{Zhu2022}. The set-up involves two flavours of electrons $\psi_\sigma(\br)$ labelled by a pseudo-spin $\sigma={\uparrow},\downarrow$ on the sphere under the influence of a magnetic monopole with $4\pi s$ charge at its centre. Due to the presence of the monopole, the single-particle eigenstates form highly degenerate quantized Landau levels~\cite{Haldane1983FQHE}. The lowest Landau level (LLL) has angular momentum $s$ and degeneracy $N_m=2s+1$. By setting the single-particle gap the leading energy scale, we project the system onto the LLL and express the second-quantized operators in terms of the creation and annihilation operators on the LLL
\begin{equation}
    \psi^\dagger_\sigma(\br)=\frac{1}{R}\sum_{m=-s}^sc^\dagger_{m\sigma} Y_{sm}^{(s)}(\br)
\end{equation}
where the single-particle wave functions are the spin-weighted spherical harmonics~\cite{Wu1976LLL}.

The spin-down and spin-up particles are respectively even and odd under the Ising $\BZ_2$ global symmetry: $\cZ c_\downarrow\cZ=c_\downarrow$ and $\cZ c_\uparrow\cZ=-c_\uparrow$. The model Hamiltonian is
\begin{equation}
    H=\frac{1}{2}\int\rd^2\br_1\,\rd^2\br_2 U(r_{12})\left(n^0(\br_1)n^0(\br_2)-n^x(\br_1)n^x(\br_2)\right)+h\int\rd^2\br\,n^z(\br)
    \label{eq:hmt_ising}
\end{equation}
where $n^i(\br)=\psi^\dagger(\br)\sigma^i\psi(\br)$ with $\psi=(\psi_\uparrow,\psi_\downarrow)^\rT$, $\sigma^i$ ($i=0,x,y,z$) are the Pauli matrices, $r_{12}=|\br_1-\br_2|$, and $U(r_{12})$ takes the form of local and super-local interactions
\begin{equation}
    U(r_{12})=g_0\delta(r_{12})+g_1\nabla^2\delta(r_{12})
    \label{eq:local_int}
\end{equation}
It can be equivalently parametrized by the pseudopotentials $U_l$. For the $n^0n^0$ term, in the orbital space,
\begin{multline}
    \int\rd^2\br_1\,\rd^2\br_2 U(r_{12})n^0(\br_1)n^0(\br_2)\\=\sum_{lm,m_1m_2m_3m_4}U_lc_{m_1\sigma}^\dagger c_{m_2\sigma'}^\dagger c_{m_3\sigma'} c_{m_4\sigma}\langle sm_1,sm_2|(2s-l)m\rangle\langle (2s-l)m|sm_3,sm_4\rangle.
\end{multline}
The interaction in the form of Eq.~\eqref{eq:local_int} corresponds to non-zero values of $U_0$, $U_1$, and $U_l=0$ for $l\geq 2$. Different from the original paper~\cite{Zhu2022}, we choose a slightly different conformal point $U_0=4.825$, $U_1=1$, and $h=3.158$ determined by tuning away the first irrelevant perturbation $\epsilon'$ in the Hamiltonian~\cite{Fardelli2024}.

The full symmetries of this model include 
\begin{itemize}[nosep]
    \item the $\rU(1)_e$ charge conservation which decouples in the infra-red,
    \item the $\SO(3)_r$ spherical rotation,
    \item the $\BZ_2$ Ising global symmetry, and 
    \item the particle-hole symmetry $\cP$ that acts as the space-time parity in the CFT in the infra-red. 
\end{itemize}
The previous exact diagonalization methods for this model make use of the Abelian and discrete subgroups to divide the Hilbert space into blocks, and diagonalize the Hamiltonian block by block. The conserved charges include the total number of electrons, the angular momentum in the $z$-direction, and the total Ising $\BZ_2$ charge 
\begin{align}
    N_e&=\sum_{m\sigma}n_{m\sigma}&L^z&=\sum_{m\sigma}mn_{m\sigma}&Z&=(-1)^{\sum_mn_{m\uparrow}}
\end{align}
where $n_{m\sigma}$ is the occupation number of each orbital. The other discrete symmetries include the particle-hole symmetry $\cP$ and the rotation $\cR_y$ by angle $\pi$ along the $y$-axis in the $L^z=0$ sector. The $L^z$ and $\cR_y$ together form an $\rO(2)$ subgroup of $\SO(3)_r$.

\subsection{Constructing the Hilbert Space}
\label{sec:method_hilbert}

The $\SO(3)$ symmetry divides the Hilbert space $\cH(N_e,m,Z)$ with a definite number of particles $N_e$, a magnetic quantum number $m$, and Ising parity $Z$ into sectors with a definite total angular momentum $l$
\begin{equation}
    \cH(N_e,m,Z)=\bigoplus_l\cH(N_e,Z,lm).
\end{equation}
To construct this space, we first decompose each occupation-number-basis state in $\cH(N_e,m,Z)$ as a direct product of two segment states containing all the spin-down and all the spin-up particles
\begin{equation}
    |N_e,m,Z\rangle=|N_{e\downarrow},m_{\downarrow}\rangle\otimes |N_{e\uparrow},m_{\uparrow}\rangle.
\end{equation}
Each two segment state carries a definite number of particles $N_{e\sigma}$ and angular momentum $m_\sigma$ in the $z$-direction. The conservation laws require $N_e=N_{e\downarrow}+N_{e\uparrow}$, $m=m_{\downarrow}+m_{\uparrow}$, and $(-1)^{N_{e\uparrow}}=Z$. The Hilbert space is thus decomposed as
\begin{equation}
    \cH(N_e,m,Z)=\bigoplus_{N_{e\downarrow}m_\downarrow,N_{e\uparrow}m_\uparrow}\cH_{\downarrow}(N_{e\downarrow},m_{\downarrow})\otimes \cH_{\uparrow}(N_{e\uparrow},m_{\uparrow}),
    \label{eq:hilb_decomp}
\end{equation}
where the sum is taken over sectors that obey the conservation laws.

The dimension of the segment Hilbert spaces $\cH_\sigma(N_{e\sigma},m_\sigma)$ is much smaller than the dimension of the full Hilbert space $\cH(N_e,m,Z)$, so the full diagonalization of the angular momentum Casimir $L^2$ in the segment Hilbert spaces $\cH_\sigma(N_{e\sigma},m_\sigma)$ is practical, allowing us to numerically divide the Hilbert space into a direct sum of spaces with definite segment total angular momentum $l_\sigma$
\begin{equation}
    \cH_\sigma(N_{e\sigma},m_\sigma)=\bigoplus_{l_\sigma}\cH_\sigma(N_{e\sigma},l_\sigma m_\sigma).
\end{equation}
We denote the state therein as $|N_{e\sigma},l_\sigma m_\sigma,\lambda_\sigma\rangle_\sigma$ where $\lambda_\sigma=1,\cdots,\dim\cH_\sigma(N_{e\sigma},l_\sigma m_\sigma)$ labels the multiplicity. 

We couple states from the two segments using Clebsch-Gordon (CG) coefficients to obtain a state in the full Hilbert space with a definite total angular momentum composition,
\begin{equation}
    |N_{e\downarrow}N_{e\uparrow}, (l_{\downarrow}l_{\uparrow})lm,\lambda_{\downarrow}\lambda_{\uparrow}\rangle=\sum_{m_\sigma}|N_{e\downarrow},l_{\downarrow} m_{\downarrow},\lambda_{\downarrow}\rangle_{\downarrow}\otimes|N_{e\uparrow},l_{\uparrow}m_{\uparrow},\lambda_{\uparrow}\rangle_{\uparrow}\langle l_{\downarrow} m_{\downarrow},l_{\uparrow} m_{\uparrow}|lm\rangle.
    \label{eq:comp_st}
\end{equation}
The total angular momentum is subject to the triangle rule $|l_{\downarrow}-l_{\uparrow}|\leq l\leq l_{\downarrow}+l_{\uparrow}$. The angular momenta $m_\sigma$ in $z$-direction of the segments are summed over and are no longer good quantum numbers. The state is labelled by the particle numbers $N_{e\downarrow},N_{e\uparrow}$, angular momentum $l_{\downarrow},l_{\uparrow}$ in each segment, the total angular momentum $l$, and angular momentum $m$ in $z$-direction, and the multiplicity $\lambda_{\downarrow},\lambda_{\uparrow}$ of each state.

We denote the Hilbert space spanned by these states as $\cH(N_{e\downarrow}N_{e\uparrow}, (l_{\downarrow} l_{\uparrow})lm)$. The desired angular-momentum-resolved Hilbert space is the direct sum
\begin{equation}
    \cH(N_e,lm,Z)=\bigoplus_{N_{e\downarrow}l_\downarrow,N_{e\uparrow}l_\uparrow}\cH(N_{e\downarrow}N_{e\uparrow}, (l_{\downarrow}l_{\uparrow})lm)
\end{equation}
over all the $N_{e\downarrow}, N_{e\uparrow}, l_{\downarrow}$, and $l_{\uparrow}$ satisfying
\begin{align}
    N_e&=N_{e\downarrow}+N_{e\uparrow}, &|l_{\downarrow}-l_{\uparrow}|&\leq l\leq l_{\downarrow}+l_{\uparrow},&(-1)^{N_{e\uparrow}}&=Z.
\end{align}

\subsection{Constructing the Operator}
\label{sec:method_operator} 

Having constructed the Hilbert space, we then discuss the construction of an operator $\Phi_{LM}:\cH(N_e,lm,Z)\to\cH(N_e',l'm',Z')$. 
The operator carries a definite electric charge, Ising parity, angular momentum $L$, and its $z$-component $M=m'-m$. More specifically, we need to evaluate the matrix element
\begin{equation}
    \langle N'_{e\downarrow}N'_{e\uparrow}, (l'_{\downarrow}l'_{\uparrow})l'm',\lambda'_{\downarrow}\lambda'_{\uparrow}|\Phi_{LM}|N_{e\downarrow}N_{e\uparrow}, (l_{\downarrow}l_{\uparrow})lm,\lambda_{\downarrow}\lambda_{\uparrow}\rangle.
\end{equation}

By the Wigner-Eckart theorem, the dependence on the magnetic quantum numbers can be extracted as a factor of a CG coefficient.
\begin{equation}
    \langle l'm'|\Phi_{LM}|lm\rangle=\langle l'm'|LM, lm\rangle\langle l'\|\Phi_L\|l\rangle
    \label{eq:wigeck}
\end{equation}
where the irrelevant indices are omitted. In the following, we will work with a component-free treatment and calculate the reduced matrix element $\langle N'_{e\downarrow}N'_{e\uparrow}, (l'_{\downarrow}l'_{\uparrow})l',\lambda'_{\downarrow}\lambda'_{\uparrow}\|\Phi_L\|N_{e\downarrow}N_{e\uparrow}, (l_{\downarrow}l_{\uparrow})l, \lambda_{\downarrow}\lambda_{\uparrow}\rangle$. We shall show that it can be expressed in terms of the reduced matrix elements of the segment operators $\langle N'_{e\sigma},l'_\sigma ,\lambda'_\sigma\|\Phi^{(\sigma,d)}_{L_\sigma }\|N_{e\sigma},l_\sigma,\lambda_\sigma\rangle_\sigma$ acting as $\Phi^{(\sigma,d)}_{L_\sigma M_\sigma}:\cH_\sigma(N_{e\sigma},l_\sigma m_\sigma)\to\cH_\sigma(N_{e\sigma}',l'_\sigma m'_\sigma)$ where $d$ specifies a coupling channel.

We decompose the operator into several channels $\Phi_L^{(d)}$ labelled by the index $d$. Each channel is a product of two segment operators $\Phi^{(\sigma,d)}_{L_\sigma}$ that act respectively on the spin-down and spin-up segments and carry definite angular momentum $L_\sigma$.
\begin{align}
    \Phi_L&=\sum_dC^{(d)}\Phi_L^{(d)},&\Phi_L^{(d)}&=[\Phi^{(\downarrow,d)}_{L_{\downarrow}}\otimes\Phi^{(\uparrow,d)}_{L_{\uparrow}}]_L\nonumber\\
    \Phi_{LM}&=\sum_dC^{(d)}\Phi_{LM}^{(d)},&\Phi_{LM}^{(d)}&=\sum_{M_{\downarrow}M_{\uparrow} }\Phi^{(\downarrow,d)}_{L_{\downarrow} M_{\downarrow}}\Phi^{(\uparrow,d)}_{L_{\uparrow} M_{\uparrow}}\langle L_{\downarrow} M_{\downarrow},L_{\uparrow} M_{\uparrow}|LM\rangle
    \label{eq:op_decomp}
\end{align}
We shall show how this decomposition is actually performed in \S~\ref{sec:method_decomp}. Hereafter, we use $[A_{L_A}\otimes B_{L_B}]_L$ to denote that the objects $A$ and $B$ carrying respective angular momenta $L_A$ and $L_B$ are coupled into total angular momentum $L$ by the rule of CG coefficients. The states can be expressed similarly by that notation 
\begin{equation}
    \|N_{e\downarrow}N_{e\uparrow}, (l_{\downarrow}l_{\uparrow})l, \lambda_{\downarrow}\lambda_{\uparrow}\rangle=\left[\|N_{e\downarrow}, l_{\downarrow}, \lambda_{\downarrow}\rangle_{\downarrow}\otimes\|N_{e\uparrow}, l_{\uparrow}, \lambda_{\uparrow}\rangle_{\uparrow}\right]_l
    \label{eq:comp_st_free}
\end{equation}
We show that the reduced matrix element $\langle N'_{e\downarrow}N'_{e\uparrow}, (l'_{\downarrow}l'_{\uparrow})l',\lambda'_{\downarrow}\lambda'_{\uparrow}\|\Phi_L^{(d)}\|N_{e\downarrow}N_{e\uparrow}, (l_{\downarrow}l_{\uparrow})l, \lambda_{\downarrow}\lambda_{\uparrow}\rangle$ can be expressed in terms of the reduced matrix elements $\langle N'_{e\sigma},l'_\sigma,\lambda'_\sigma\|\Phi^{(\sigma,d)}_{L_\sigma}\|N_{e\sigma},l_\sigma,\lambda_\sigma\rangle$ of the segment operators. Similar to Eq.~\eqref{eq:wigeck}, it is related to the full element by
\begin{equation}
    \langle N'_{e\sigma},l'_\sigma m'_\sigma,\lambda'_\sigma|\Phi^{(\sigma,d)}_{L_\sigma M_\sigma}|N_{e\sigma},l_\sigma m_\sigma,\lambda_\sigma\rangle_\sigma=\langle l'_\sigma m'_\sigma|L_\sigma M_\sigma,l_\sigma m_\sigma\rangle\langle N'_{e\sigma},l'_\sigma,\lambda'_\sigma\|\Phi^{(\sigma,d)}_{L_\sigma}\|N_{e\sigma},l_\sigma,\lambda_\sigma\rangle_\sigma
    \label{eq:op_red_seg}
\end{equation}

The matrix element of the composite operator is written in terms of the matrix elements of the segment operators through an angular momentum recoupling
\begin{align}
    &\hphantom{{}={}}\langle N'_{e\downarrow}N'_{e\uparrow}, (l'_{\downarrow}l'_{\uparrow})l',\lambda'_{\downarrow}\lambda'_{\uparrow}\|\Phi_L^{(d)}\|N_{e\downarrow}N_{e\uparrow}, (l_{\downarrow}l_{\uparrow})l,\lambda_{\downarrow}\lambda_{\uparrow}\rangle\nonumber\\
    &=\left[{}_{\downarrow}\langle N'_{e\downarrow},l'_{\downarrow},\lambda'_{\downarrow}\|\otimes{}_{\uparrow}\langle N'_{e\uparrow},l'_{\uparrow},\lambda'_{\uparrow}\|\right]_{l'}\left[\Phi^{(\downarrow,d)}_{L_{\downarrow}}\otimes\Phi^{(\uparrow,d)}_{L_{\uparrow}}\right]_L\left[\|N_{e\downarrow},l_{\downarrow},\lambda_{\downarrow}\rangle_{\downarrow}\otimes\|N_{e\uparrow},l_{\uparrow},\lambda_{\uparrow}\rangle_{\uparrow}\right]_l\nonumber\\
    &=\langle N'_{e\downarrow},l'_{\downarrow},\lambda'_{\downarrow}\|\Phi^{(\downarrow,d)}_{L_{\downarrow}}\| N_{e\downarrow},l_{\downarrow},\lambda_{\downarrow}\rangle_{\downarrow}\langle N'_{e\uparrow},l'_{\uparrow},\lambda'_{\uparrow}\|\Phi^{(\uparrow,d)}_{L_{\uparrow}}\|N_{e\uparrow},l_{\uparrow},\lambda_{\uparrow}\rangle_{\uparrow}\nonumber\\
    &\qquad\qquad\times(-1)^{N_{e\downarrow}(N'_{e\uparrow}-N_{e\uparrow})}\sqrt{(2l'_{\downarrow}+1)(2l'_{\uparrow}+1)(2l+1)(2L+1)}\begin{Bmatrix}
        l_{\downarrow}&l_{\uparrow}&l\\
        L_{\downarrow}&L_{\uparrow}&L\\
        l'_{\downarrow}&l'_{\uparrow}&l'
    \end{Bmatrix}
    \label{eq:op_comp}
\end{align}
Here the first line represents the angular momentum composition in the channel $((l_{\downarrow}l_{\uparrow})l,(L_{\downarrow}L_{\uparrow})L)l'$ --- first coupling $l_{\downarrow}$ and $l_{\uparrow}$ to $l$ and coupling $L_{\downarrow}$ and $L_{\uparrow}$ to $L$, and then composing $l$ and $L$ into $l'$ --- the second line represents recoupling it in the channel $((l_{\downarrow}L_{\downarrow})l'_{\downarrow},(l_{\uparrow}L_{\uparrow})l'_{\uparrow})l'$, their inner product is specified by the Wigner $9j$-symbol
\begin{multline}
    \langle((l_{\downarrow}l_{\uparrow})l,(L_{\downarrow}L_{\uparrow})L)l'|((l_{\downarrow}L_{\downarrow})l'_{\downarrow},(l_{\uparrow}L_{\uparrow})l'_{\uparrow})l'\rangle\\=\sqrt{(2l'_{\downarrow}+1)(2l'_{\uparrow}+1)(2l+1)(2L+1)}\begin{Bmatrix}
        l_{\downarrow}&l_{\uparrow}&l\\
        L_{\downarrow}&L_{\uparrow}&L\\
        l'_{\downarrow}&l'_{\uparrow}&l'
    \end{Bmatrix}.
\end{multline}
An additional sign factor comes from permuting the fermionic operators.

This completes the decomposition of the composite-operator matrix elements in terms of the segment-operator matrix elements. In practice, the segment operators require relatively small memory. They are calculated and stored prior to the diagonalization. The reduced matrix elements are calculated using Eq.~\eqref{eq:op_red_seg} with representative values of $m'_\sigma, M_\sigma$, and $m_\sigma$, usually taken as $0$ or $1/2$ (depending on the angular momentum being $\BZ$ or $\BZ+1/2$), or $m'_\sigma=M_\sigma=1$, $m_\sigma=0$ when the CG coefficient vanishes at $l'_\sigma,L_\sigma,l_\sigma\in\BZ$ and $l'_\sigma+L_\sigma+l_\sigma\in2\BZ+1$. 

\subsection{Decomposing the Hamiltonian}
\label{sec:method_decomp} 

In this section, we demonstrate the decomposition~\eqref{eq:op_decomp} with the example of the Hamiltonian~\eqref{eq:hmt_ising} of the Ising model. In the orbital space, the Hamiltonian is
\begin{align}
    H&=U_0H_{U,0}+U_1H_{U,1}+hH_h\nonumber\\
    H_{U,0}&=\sum_{mm_1m_2m_3m_4}c_{m_1\downarrow}^\dagger c_{m_2\uparrow}^\dagger c_{m_3\uparrow}c_{m_4\downarrow}\langle sm_1,sm_2|(2s)m\rangle\langle (2s)m|sm_3,sm_4\rangle\nonumber\\
    H_{U,1}&=\sum_{mm_1m_2m_3m_4}\frac{1}{2}\langle sm_1,sm_2|(2s-1)m\rangle\langle (2s-1)m|sm_3,sm_4\rangle\left(c_{m_1\downarrow}^\dagger c_{m_2\downarrow}^\dagger c_{m_3\uparrow}c_{m_4\uparrow}\right.\nonumber\\
    &\qquad\qquad\left.+c_{m_1\uparrow}^\dagger c_{m_2\uparrow}^\dagger c_{m_3\downarrow} c_{m_4\downarrow}
    +c_{m_1\downarrow}^\dagger c_{m_2\downarrow}^\dagger c_{m_3\downarrow}c_{m_4\downarrow}+c_{m_1\uparrow}^\dagger c_{m_2\uparrow}^\dagger c_{m_3\uparrow}c_{m_4\uparrow}\right)\nonumber\\
    H_h&=\sum_m(c_{m\uparrow}^\dagger c_{m\uparrow}-c_{m\downarrow}^\dagger c_{m\downarrow})
\end{align}
Simplifications are made using the symmetry property of the CG coefficients. The polarization term and the $U_1$ term decompose straightforwardly
\begin{align}
    H_h&=N_m\left[\BI^{(\downarrow)}_0\otimes n^{(\uparrow)}_0\right]_0-N_m\left[n^{(\downarrow)}_0\otimes\BI^{(\uparrow)}_{0}\right]_0\nonumber\\
    H_{U,1}&=\frac{1}{2}\left(\left[(\Delta^\dagger)^{(\downarrow)}_{2s-1}\otimes(\Delta)^{(\uparrow)}_{2s-1}\right]_0+\left[(\Delta)^{(\downarrow)}_{2s-1}\otimes(\Delta^\dagger)^{(\uparrow)}_{2s-1}\right]_0\right.\nonumber\\
    &\qquad\qquad\left.+\left[(H_1)^{(\downarrow)}_0\otimes\BI_0^{(\uparrow)}\right]_0+\left[\BI^{(\downarrow)}_0\otimes (H_1)_0^{(\uparrow)}\right]_0\right).
    \label{eq:decomp_hU1}
\end{align}
where
\begin{align*}
    (H_1)^{(\sigma)}_{00}&=\sum_{mm_1m_2m_3m_4}c_{m_1\sigma}^\dagger c_{m_2\sigma}^\dagger c_{m_3\sigma} c_{m_4\sigma}\langle sm_1,sm_2|(2s-1)m\rangle\langle (2s-1)m|sm_3,sm_4\rangle\nonumber\\
    (\Delta^\dagger)^{(\sigma)}_{lm}&=\sum_{m_1m_2}c^\dagger_{m_1\sigma} c^\dagger_{m_2\sigma}\langle lm|sm_1,sm_2\rangle\\
    n_{lm}^{(\sigma)}&=\sum_{m_1m_2}c_{m_1\sigma}^\dagger c_{m_2\sigma}\frac{(-1)^{m_2-s}}{\sqrt{2s+1}}\langle lm|sm_1,s(-m_2)\rangle
\end{align*}

For the $U_0$ term, we use the pseudopotential recoupling: the pseudopotential corresponds to the coupling in the pairing channel
\begin{align}
    \sum_l\tilde{U}_{l}\left[(c_{1,\downarrow}^\dagger c_{2,\uparrow}^\dagger)_{l}\otimes(c_{3,\downarrow}c_{4,\uparrow})_{l}\right]_0
\end{align}
where the operators are composed using Clebsch-Gordon (CG) coefficients, and the only non-zero term is $\tilde{U}_{2s}=U_0/\sqrt{4s+1}$. The fermions are labelled 1, 2, 3, and 4 for convenience. One needs to recouple it into the density channel
\begin{align}
    \sum_l\tilde{U}_{l}\left[(c_{1,\downarrow}^\dagger c_{2,\uparrow}^\dagger)_{l}\otimes(c_{3,\uparrow}c_{4,\downarrow})_{l}\right]_0=\sum_j\tilde{V}_j\left[(c_{1,\downarrow}^\dagger c_{4,\downarrow})_{j}\otimes(c_{2,\uparrow}^\dagger c_{3,\uparrow})_j\right]_0
\end{align}
The relation of $\tilde{U}$ and $\tilde{V}$ can be expressed in a Dirac bra-ket notation, which corresponds to a $9j$-symbol\footnote{The $9j$-symbol reduces to a $6j$-symbol when one element is zero
\begin{equation*}
    \sum_l\tilde{U}_l\,(-1)^{s_2+s_3+2s_4+j}\sqrt{(2l+1)(2j+1)}\begin{Bmatrix}
        s_1&s_2&l\\s_3&s_4&j
    \end{Bmatrix}.
\end{equation*}}
\begin{align}
    \tilde{V}_j&=\sum_l\tilde{U}_l\langle((s_1s_2)l,(s_3s_4)l)0|((s_1s_4)j,(s_2s_3)j)0\rangle\nonumber\\
    &=\sum_l\tilde{U}_l\,(-1)^{s_3+s_4-l}(2l+1)(2j+1)\begin{Bmatrix}s_1&s_2&l\\s_4&s_3&l\\j&j&0\end{Bmatrix}.
\end{align}
The sign comes from exchanging $s_3$ and $s_4$, and $s_1=s_2=s_3=s_4=s$. Hence 
\begin{align}
    H_{U,0}&=2\sum_j\tilde{V}_jN_m\left[n^{(\downarrow)}_j\otimes n^{(\uparrow)}_j\right]_0,&
    \tilde{V}_j&=(2j+1)\sqrt{4s+1}\begin{Bmatrix}s&s&2s\\s&s&2s\\j&j&0\end{Bmatrix}.
    \label{eq:decomp_U0}
\end{align}

Eqs.~\eqref{eq:decomp_hU1} and \eqref{eq:decomp_U0} complete the decomposition of the Hamiltonian into $(N_m+6)$ channels of direct products of operators that act on a single segment. 

\subsection{Numerical Procedure}
\label{sec:method_summary} 

We now describe the steps for computing the lowest-lying eigenstates of the fuzzy-sphere Ising model in a sector with fixed total $\SO(3)$ angular momentum $l$, its $z$-component $m$, and Ising parity $Z$. Without loss of generality, we take even $N_m$, $m=0$ and $Z=+1$ as an example.

\paragraph{Build the segment spaces} First we generate $\cH_\sigma(N_{e\sigma},m_\sigma)$ in the occupation-number basis where $\sigma={\uparrow},\downarrow$, $N_{e\sigma}=0,2,4,\dots,N_m$, and $m_\sigma=0$. By diagonalizing the full matrix of $L^2_\sigma$, we resolve the total angular momentum. We store all the states $|N_{e\sigma},l_\sigma m_\sigma,\lambda_\sigma\rangle_\sigma$ in the occupation-number basis.

\paragraph{Build the composite space} We compose the segment states into the composite state $|N_{e\downarrow}N_{e\uparrow},(l_{\downarrow}l_{\uparrow})lm,\lambda_{\downarrow}\lambda_{\uparrow}\rangle$ using Eq.~\eqref{eq:comp_st} or equivalently Eq.~\eqref{eq:comp_st_free}. In this process, we record the Hilbert-space dimension and the list of $(N_{e\downarrow}N_{e\uparrow},l_{\downarrow}l_{\uparrow},\lambda_{\downarrow}\lambda_{\uparrow})$ for each composite state.

\paragraph{Decompose the Hamiltonian} We write the Hamiltonian in the form of Eq.~\eqref{eq:op_decomp}
\begin{equation}
    H=\sum_d c^{(d)}\left[H_{L_{\downarrow}}^{(\downarrow,d)}\otimes H_{L_{\uparrow}}^{(\uparrow,d)}\right]_0
\end{equation}
where $H_{L_\sigma}^{(\sigma,d)}$ acts on a single segment. This decomposition is elaborated in \S~\ref{sec:method_decomp}. 

\paragraph{Build the segment operators} We calculate and store the reduced matrix elements of the segment operators as a collection of matrices, one for each sector 
\begin{equation}
    \Big((\hat{H}_{L_\sigma}^{(\sigma,d)})_{N'_{e\sigma}l'_\sigma}^{N_{e\sigma}l_\sigma}\Big)_{\lambda'_\sigma}\Big.^{\lambda_\sigma}=\langle N'_{e\sigma},l'_\sigma,\lambda'_\sigma\|H^{(\sigma,d)}_{L_\sigma}\| N_{e\sigma},l_\sigma,\lambda_\sigma\rangle_\sigma.
\end{equation}
The reduced matrix elements are derived from their components by Eq.~\eqref{eq:op_red_seg} taking $m'_\sigma,M_\sigma,m_\sigma=0$ or $1/2$. In the case of vanishing CG coefficient at $l'_\sigma,L_\sigma,l_\sigma\in\BZ$, $l'_\sigma+L_\sigma+l_\sigma\in 2\BZ+1$, we take instead $m_\sigma=1,M_\sigma=-1$. The state is generated by applying the ladder operator $|N_{e\sigma},l_\sigma1,\lambda_\sigma\rangle_\sigma=L^+|N_{e\sigma},l_\sigma0,\lambda_\sigma\rangle_\sigma/\sqrt{l_\sigma(l_\sigma+1)}$.

\paragraph{Build the composite operators} The segment operators are composed into the composite operator by Eq.~\eqref{eq:op_comp}. The full matrix of the composite operator is never materialized, although the $9j$-symbols for each sector
\begin{equation}
    C^{N'_{e\downarrow}N'_{e\uparrow}, (l'_{\downarrow}l'_{\uparrow})l}_{N_{e\downarrow}N_{e\uparrow}, (l_{\downarrow}l_{\uparrow})l}=(-1)^{N_{e\downarrow}(N'_{e\uparrow}-N_{e\uparrow})}\sqrt{(2l'_{\downarrow}+1)(2l'_{\uparrow}+1)(2l+1)(2L+1)}\begin{Bmatrix}
        l_{\downarrow}&l_{\uparrow}&l\\
        L_{\downarrow}&L_{\uparrow}&L\\
        l'_{\downarrow}&l'_{\uparrow}&l
    \end{Bmatrix}
\end{equation}
are stored for acceleration. 

\paragraph{Finding the eigenstates through Lanczos algorithm} Finally, the leading eigenstates are obtained through the Lanczos algorithm. The input to the algorithm is a mapping in the Hilbert space $\cH(N_e,lm,Z)$ from a given state  $|\Psi\rangle$ to the product $H|\Psi\rangle$ of Hamiltonian and that state. Since the full matrix is not materialized, to take this product, one goes through all the decomposition channels $d$; for each channel, one goes through all the initial sectors $N_{e\downarrow}N_{e\uparrow},l_{\downarrow}l_{\uparrow}$ and all the final sectors $N_{e\downarrow}N_{e\uparrow},l_{\downarrow}l_{\uparrow}$ that satisfy the conservation law and the triangular rule of the angular momenta. The elements within can be expressed as a matrix product
\begin{align}
    |\Psi\rangle&=\sum_{N_{e\downarrow}N_{e\uparrow}, l_{\downarrow}l_{\uparrow},\lambda_{\downarrow}\lambda_{\uparrow}}(\hat{\psi}_{N_{e\downarrow}N_{e\uparrow},l_{\downarrow}l_{\uparrow}})_{\lambda_{\downarrow}\lambda_{\uparrow}}\|N_{e\downarrow}N_{e\uparrow}, (l_{\downarrow}l_{\uparrow})l,\lambda_{\downarrow}\lambda_{\uparrow}\rangle\label{eq:st_comp}\\
    H|\Psi\rangle&=\sum_{N'_{e\downarrow}N'_{e\uparrow}, l'_{\downarrow}l'_{\uparrow},\lambda'_{\downarrow}\lambda'_{\uparrow}}(\hat{\psi}'_{N'_{e\downarrow}N'_{e\uparrow},l'_{\downarrow}l'_{\uparrow}})_{\lambda'_{\downarrow}\lambda'_{\uparrow}}\|N'_{e\downarrow}N'_{e\uparrow}, (l'_{\downarrow}l'_{\uparrow})l,\lambda'_{\downarrow}\lambda'_{\uparrow}\rangle\nonumber\\
    \hat{\psi}'_{N'_{e\downarrow}N'_{e\uparrow},l'_{\downarrow}l'_{\uparrow}}&=\sum_{d,N_{e\downarrow}N_{e\uparrow}l_{\downarrow}l_{\uparrow}}c^{(d)}C^{N'_{e\downarrow}N'_{e\uparrow}, (l'_{\downarrow}l'_{\uparrow})l}_{N_{e\downarrow}N_{e\uparrow}, (l_{\downarrow}l_{\uparrow})l}(\hat{H}_{L_{\downarrow}}^{(\downarrow,d)})_{N'_{e\downarrow}l'_{\downarrow}}^{N_{e\downarrow}l_{\downarrow}}\hat{\psi}_{N_{e\downarrow}N_{e\uparrow},l_{\downarrow}l_{\uparrow}}\left((\hat{H}_{L_{\uparrow}}^{(\uparrow,d)})_{N'_{e\uparrow}l'_{\uparrow}}^{N_{e\uparrow}l_{\uparrow}}\right)^\rT
\end{align}
where the hat denotes that the elements are arranged as a matrix. Briefly speaking, the Lanczos algorithm is an iterative method to extract the extremal eigenvalues and eigenvectors of a Hermitian linear operator acting on a large Hilbert space. Starting with an initial vector, each iteration enlarges an orthonormal basis of the Krylov subspace and obtains increasingly accurate approximations to the desired eigenvectors within this subspace. The Krylov subspace of dimension $r$ is generated by repeatedly applying $H$ to an initial vector $|i\rangle$
\begin{equation}
    \mathcal{K}_r(H,|i\rangle)=\operatorname{span}\left\{|i\rangle,H|i\rangle,H^2|i\rangle,\dots,H^{r-1}|i\rangle\right\}.
\end{equation}

\subsection{Particle-Hole Symmetry}
\label{sec:method_ph}

We next discuss the implementation of the particle-hole symmetry $\cP$. It acts on the single-particle operators as $\cP c_{m\downarrow}\cP=(-1)^{s+m}c_{(-m)\uparrow}^\dagger$ and $\cP c_{m\uparrow}\cP=-(-1)^{s+ m}c_{(-m)\downarrow}^\dagger$, and therefore acts on the segment space as
\begin{equation}
    \cP:\cH_{\downarrow}(N_{e\downarrow},l_{\downarrow}m_{\downarrow})\to\cH_{\uparrow}(N_e-N_{e\downarrow}, l_{\downarrow}m_{\downarrow}).
\end{equation}
We build the $\cH_{\uparrow}$ spaces by acting particle-hole transformation on the $\cH_{\downarrow}$ spaces
\begin{equation}
    \cP|N_{e\downarrow},l_{\downarrow}m_{\downarrow},\lambda_{\downarrow}\rangle_{\downarrow}=|N_e-N_{e\downarrow}, l_{\downarrow}m_{\downarrow},\lambda_{\downarrow}\rangle_{\uparrow}\in\cH_{\uparrow}(N_e-N_{e\downarrow}, l_{\downarrow}m_{\downarrow}).
\end{equation}
Conversely 
\begin{equation}
    \cP|N_{e\uparrow}, l_{\uparrow}m_{\uparrow},\lambda_{\uparrow}\rangle_{\uparrow}=(-1)^{N_{e\uparrow}}|N_e-N_{e\uparrow},l_{\uparrow}m_{\uparrow},\lambda_{\uparrow}\rangle_{\downarrow}
\end{equation}
Therefore the particle-hole symmetry acts on the composite space as 
\begin{equation}
    \cP|N_{e\downarrow}N_{e\uparrow},(l_{\downarrow}l_{\uparrow})l,\lambda_{\downarrow}\lambda_{\uparrow}\rangle=Z(-1)^{l_{\downarrow}+l_{\uparrow}-l}|N_{e\downarrow}N_{e\uparrow},(l_{\uparrow}l_{\downarrow})l,\lambda_{\uparrow}\lambda_{\downarrow}\rangle
\end{equation}
where we have used $N_{e\downarrow}+N_{e\uparrow}=N_e$ and $(-1)^{N_{e\uparrow}}=Z$, and a sign factor comes from the symmetry property of the CG coefficient. Therefore, a state in the form of Eq.~\eqref{eq:st_comp} with parity $P=\pm 1$ under particle-hole symmetry must obey
\begin{equation}
    (\hat{\psi}_{N_{e\downarrow}N_{e\uparrow},l_{\downarrow}l_{\uparrow}})_{\lambda_{\downarrow}\lambda_{\uparrow}}=PZ(-1)^{l_{\downarrow}+l_{\uparrow}-l}(\hat{\psi}_{N_{e\downarrow}N_{e\uparrow},l_{\uparrow}l_{\downarrow}})_{\lambda_{\uparrow}\lambda_{\downarrow}}.
\end{equation}
In practice, we obtain the parity-odd eigenstates by projecting the state to the parity-odd space every time after acting the Hamiltonian, \emph{i.~e.}~we feed $|\Psi\rangle\mapsto\frac{1}{2}(1-\cP)H|\Psi\rangle$ as the linear mapping into the Lanczos algorithm.

\subsection{Generalization to Other Models}
\label{sec:method_other}

The $\SO(3)$-resolving exact diagonalization could be applied to other models on the fuzzy sphere, such as the $\SO(5)$ deconfined criticality~\cite{Zhou2023}, the $\Sp(N)$-symmetric CFTs~\cite{Zhou2024Oct}, $\SU(2)_1$-Higgs theory~\cite{Zhou2025Jul}, and the free Majorana fermion~\cite{Zhou2025Sep}. To accommodate other models, the algorithm could be generalized in several ways.

\begin{enumerate}[nosep, wide]
    \item The Hilbert space could be partitioned into more than two segments. In this case, the angular momentum composes successively along a chain, so a state is specified not only by the total angular momentum $l$, but also by the cumulative angular momentum of parts $1\cdots p$ for each $p$
    \begin{equation}
        \|\{l\}\rangle=\left\|\left(\cdots\left.\left((l_1 l_2)l_{12}\,l_3\right)l_{123}\cdots l_p\right)l_{1\cdots p}\cdots l_{N_p}\right)l\right\rangle
    \end{equation}
    where irrelevant indices are omitted. Here we use $\{l\}$ as a short-hand notation for the coupling channel. Each channel is specified by $2N_p-1$ angular momenta in total --- $l_p$ ($p=1,\dots,N_p$), $l_{1\cdots p}$ ($p=2,\dots,N_p$), where $N_p$ is the number of parts, and $l_{1\cdots N_p}=l$. Its components are 
    \begin{equation}
        |\{l\}m\rangle=\sum_{\{m_p\},\{m_{1\cdots p}\}}\bigotimes_{p=1}^{N_p}|l_pm_p\rangle_p\prod_{p=2}^{N_p}\langle l_{1\cdots(p-1)}m_{1\cdots(p-1)},l_pm_p|l_{1\cdots p}m_{1\cdots p}\rangle
    \end{equation}
    where $m_{1\cdots N_p}=m$. The coupling of segment operators is specified in a similar way. 
    \begin{align}
        \Phi&=\sum_{\{L\}}c_{\{L\}}\Phi_{\{L\}}&\nonumber\\
        \Phi_{\{L\}}&=\Big[\dots[(\Phi_1)_{L_1}\otimes(\Phi_2)_{L_2}]_{L_{12}}\dots (\Phi_p)_{L_p}\big]_{L_{1\cdots p}} \cdots (\Phi_{N_p})_{L_{N_p}}\Big]_L
    \end{align}
    Calculating the reduced matrix element involves $(N_p-1)$ angular-momentum recouplings. Each recoupling produces a $9j$-symbol, so
    \begin{multline}
        \langle \{l'\}\|\Phi_{\{L\}}\|\{l\}\rangle
        =\prod_{p=1}^{N_p}\langle l'_p\|\Phi_{p,L_p}\|l_p\rangle\prod_{p<p'}(-1)^{F_p(F'_{p'}-F_{p'})}\\
        \times\prod_{p=2}^{N_p}\left[\sqrt{(2l_{1\cdots p}+1)(2L_{1\cdots p}+1)(2l'_{1\cdots (p-1)}+1)(2l'_p+1)}
        \begin{Bmatrix}l_{1\cdots(p-1)}&l_p&l_{1\cdots p}\\L_{1\cdots(p-1)}&L_p&L_{1\cdots p}\\l'_{1\cdots(p-1)}&l'_p&l'_{1\cdots p}\end{Bmatrix}\right]
    \end{multline}
    where $F_p=0,1$ is the fermion parity in the segment $p$.
    \item Besides the $\SO(3)$ rotation, non-Abelian flavour symmetries that act within an individual segment could be resolved by diagonalizing their quadratic Casimirs $C_{2,p}$ simultaneously with the total angular momentum $L^2_p$. The number of particles could also be generalized into a set of diagonal quantum numbers $\mathbf{Q}$. Altogether, a state within the Hilbert space is labelled by the angular-momentum coupling channel $\{l\}$, the diagonal quantum numbers $\mathbf{Q}_p$, the quadratic Casimir $C_{2,p}$, and the multiplicity $\lambda_p$ within each part 
    \begin{equation*}
        \left\|\{\mathbf{Q}_p\},\{l\},\{C_{2,p}\},\{\lambda_p\}\right\rangle.
    \end{equation*} 
\end{enumerate}

We analyse several concrete examples. 

\paragraph{The $\SO(5)$ deconfined criticality} The model~\cite{Zhou2023} contains four flavours of fermions with a global symmetry $\SO(5)\equiv\Sp(2)/\BZ_2$. It is partitioned into two segments, each containing two flavours with a $\SU(2)$ symmetry. The total implemented global symmetry is 
\begin{equation*}
    \rO(4)\equiv \frac{\SU(2)_1\times\SU(2)_2}{\BZ_2^C}\rtimes \BZ_2^\text{exch.}\subset \SO(5)
\end{equation*}
where $\BZ_2^C$ is the common centre of the two $\SU(2)$ factors. Taking the quotient constrains $s_1+s_2\in\BZ$. The improper $\BZ_2^\text{exch.}$ acts as exchanging the two segments. The particle-hole symmetry could also be implemented. 

\paragraph{The $\Sp(N)$ CFT} The model~\cite{Zhou2024Oct} contains $2N$ flavours of fermions at filling $2M$ with a global symmetry $\Sp(N)/\BZ_2$. It is partitioned into $N$ segments, each containing two flavours with a $\SU(2)$ symmetry. The total implemented global symmetry is $\SU(2)_1\times\SU(2)_2\times\cdots\times\SU(2)_N\subset\Sp(N)$ up to $\BZ_2$ quotients.

\paragraph{The $\SU(2)_1$-Higgs Chern-Simons-matter theory} The model~\cite{Zhou2025Jul} contains two flavours of fermions and one flavour of bosons. It is partitioned into two segments of respectively fermions and bosons. The $\SO(3)$ global symmetry that acts within the fermion segment is resolved. 

\paragraph{The free Majorana fermion} The model~\cite{Zhou2025Sep} contains one flavour of fermions and one flavour of bosons. It is partitioned into two segments of respectively fermions and bosons. 

\subsection{Performance Benchmark}
\label{sec:method_benchmark}

We benchmark our $\SO(3)$-resolving algorithm by calculating the 10 lowest-lying states in the spin-0 Ising-even sector. Our code is publicly available in the module \texttt{SO3lver} of the package \texttt{FuzzifiED}~\cite{FuzzifiED}. We perform the calculation on 2.4 GHz Intel\textsuperscript{\textregistered} Xeon\textsuperscript{\textregistered} Gold 6148 CPU with 40 cores and maximal RSS memory of 180 GiB. We compare the performance with the $\rU(1)$-resolving calculation of 10 states in the $l^z=0$ spin-even, particle-hole-even, and Ising-even sector.

The maximal system size we reach is $N_m=22$ with a total memory usage of 101 GiB, of which $2/3$ is allocated to the segment operators composing the Hamiltonian. The remainder is allocated to the Hilbert space, the Krylov space, and the computational overhead. In comparison, the $\rU(1)$-resolving calculation can reach only $N_m=18$. For $N_m=18$, the $\SO(3)$-resolving ED reduces the Hilbert-space dimension by a factor of 130, the Hamiltonian storage by a factor of 120, the peak memory usage by a factor of 30, the run time for diagonalization by a factor of 140, and the total run time by a factor of 70 compared with the $\rU(1)$-resolving ED (Table~\ref{tbl:bench_ising}).

\newcommand{\ee}[1]{\times 10^{#1}}
\begin{table}[htbp]
    \centering
    \small
    \setlength{\tabcolsep}{3pt}
    \begin{tabular}{r|rrrrrr}
        \hline\hline
        $N_m$ & 22 & 20 & 18 & 16 & 14 & 12\\
        \hline 
        &\multicolumn{6}{c}{$\SO(3)$-resolving ED}\\
        \hline
        Hilbert-space dimension & $2.07\ee{7}$ & $2.07\ee{6}$ & $2.18\ee{5}$ & $2.46\ee{4}$ & $3064$ & $448$\\
        Hilbert-space memory & 10.6 GiB & 938 MiB & 84.6 MiB & 8.09 MiB & 886 KiB & 126 KiB\\
        Hamiltonian memory & 68.7 GiB & 5.38 GiB & 465 MiB & 42.4 MiB & 4.87 MiB & 879 KiB\\
        Peak memory & 101 GiB & 11.7 GiB & 2.85 GiB & 1.05 GiB & 907 MiB & 839 MiB \\
        Time to build Hilbert space & 17.0 min & 28.8 s & 2.46 s & 0.36 s & 67 ms & 16 ms\\
        Time to build Hamiltonian & 27.8 min & 70.9 s & 6.76 s & 1.32 s & 0.14 s & 61 ms\\
        Diagonalization time & 49.6 min & 104 s & 9.10 s & 1.96 s & 0.83 s & 0.38 s\\
        Total time & 94.4 min & 209 s & 21.7 s & 6.02 s & 2.07 s & 0.91 s\\
        \hline
        &\multicolumn{6}{c}{$\rU(1)$-resolving ED}\\
        \hline
        Hilbert-space dimension & & & $2.83\ee{7}$ & $2.24\ee{6}$ & $1.84\ee{5}$ & $1.61\ee{4}$\\
        Hilbert-space memory & & & 8.65 GiB & 700 MiB & 57.1 MiB & 4.91 MiB \\
        Hamiltonian memory & & & 55.2 GiB & 3.11 GiB & 179 MiB & 9.99 MiB\\
        Peak memory & & & 77.8 GiB & 5.27 GiB & 776 MiB & 475 MiB \\
        Time to build Hilbert space & & & 116 s & 7.84 s & 0.61 s & 80 ms\\
        Time to build Hamiltonian & & & 148 s & 7.98 s & 0.57 s & 31 ms\\
        Diagonalization time  & & & 21.6 min & 73.3 s & 4.26 s & 0.27 s\\
        Total time & & & 25.9 min & 89.1 s & 5.44 s & 0.39 s \\
        \hline\hline
    \end{tabular}
    \caption{Statistics on the Hilbert-space dimensions, memory usage and the time cost for the $\SO(3)$- and $\rU(1)$-resolving ED at various system sizes $N_m$. The total memory and time include the overhead. For the $\SO(3)$-resolving ED, the 10 lowest-lying states with $l=0$ and Ising-even are calculated; for the $\rU(1)$-resolving ED, the 10 lowest-lying states with $l^z=0$, spin-even, particle-hole-even and Ising-even are calculated. The timings exclude just-in-time compilation. }
    \label{tbl:bench_ising}
\end{table}

\begin{figure}[htbp]
    \centering
    \hfill
    \begin{minipage}[c]{0.5\linewidth}
        \includegraphics[width=\linewidth]{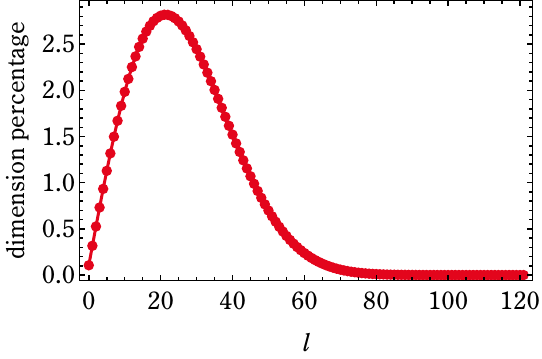}
    \end{minipage}
    \hfill
    \small
    \begin{minipage}[c]{0.36\linewidth}
        \begin{tabular}{rr|rr}
            \hline\hline
            $l$ & $\dim\cH$ & $l$ & $\dim\cH$ \\
            \hline
            $0$ & $2.06\ee{7}$ & $5$ & $2.20\ee{8}$\\
            $1$ & $6.18\ee{7}$ & $6$ & $2.57\ee{8}$\\
            $2$ & $1.02\ee{8}$ & $7$ & $2.92\ee{8}$\\
            $3$ & $1.42\ee{8}$ & $8$ & $3.25\ee{8}$\\
            $4$ & $1.82\ee{8}$ & $9$ & $3.57\ee{8}$\\
            \hline\hline
        \end{tabular}\\
        \vspace{\baselineskip}
    \end{minipage}
    \hfill{}
    \caption{The dimensions of spin-$l$ sectors of the Ising-even Hilbert space. (Left) The proportion in the Ising-even $l^z=0$ Hilbert space; (Right) the dimensions of $l<10$ spaces. }
    \label{fig:dim_l}
\end{figure}

For the spinning sectors with $l>0$, although the Hilbert space grows substantially --- the spin-1 and spin-5 sectors have approximately 3 and 10 times the dimension of the spin-0 sector (Figure~\ref{fig:dim_l}) --- the growth of the storage required for the Hilbert space and the Hamiltonian is marginal, since most space is devoted to the segment spaces and the segment operators, which does not depend on the spin.

For the other models, we list the dimension of the Hilbert space, the memory for the Hilbert space, and the memory for the Hamiltonian for these models using different methods in Table~\ref{tbl:bench_other}. We consider the sector of the $\SO(3)$ spin-0, flavour-singlet sector. The $\SO(3)$-resolving method reduces the storage memory by a factor of 170 and 25 for the $\SO(5)$ deconfined criticality and $\Sp(3)$ CFT respectively at the same system size, and increases the maximal system size by 2 at comparable storage memory. For the models involving bosons, due to large size of the segment space of the bosons, the $\SO(3)$-resolving method does not reduce the memory usage, but it could reduce the Hilbert-space dimension by a factor of 200 and 160, and could help resolve more states within certain $\SO(3)$-rotation spin and $\SO(3)$-flavour spin. 

On the other hand, as we will discuss in detail in \S~\ref{sec:results_trunc}, the CFT states of $\SU(2)_1$-Higgs transition mainly live in the zero-energy space of the parent Hamiltonian $H_\text{FQH}=[b^\dagger b^\dagger]_{2s_b}\cdot[bb]_{2s_b}$ of the $1/2$-Laughlin wave function, \textit{i.~e.}~the states whose boson sectors fall within the quasi-hole space of the Laughlin wave function. This allows the projection of the Hilbert space of the boson segment into the quasi-hole space, practically generated by squeezing the $(1,2)$-admissible Jack roots~\cite{Bernevig:2007nek,Bernevig:2008rda,Bernevig2008,Bernevig2009}.\footnote{The code for generating Jack states is publicly available in the module \texttt{JackToolkit} of the package \texttt{FuzzifiED}~\cite{FuzzifiED}.} Similarly, the CFT states of the free Majorana fermion mainly live in the zero-energy space of the parent Hamiltonian $H_\text{(bPf)}=[b^\dagger b^\dagger b^\dagger]_{3s_b}\cdot[bbb]_{3s_b}$ of the bosonic Pfaffian state generated by squeezing the $(2,2)$-admissible Jack roots. The quasi-hole-space projection reduces the Hilbert-space dimension by a factor of 2000 and 1000 respectively, reduces the memory usage by a factor of 370 and 150 compared with the $\rU(1)$-resolving ED at the same system size, and increases the maximal system size by 3 at comparable storage memory. 

\begin{table}[htbp]
    \centering
    \begin{tabular}{llr|rrr}
        \hline\hline
        Model & Method & $N_m$ & Hilbert-space & Hilbert-space & Hamiltonian \\[-4pt]
        & & & dimension & memory & memory \\ 
        \hline
        $\SO(5)$ DQCP & $\rU(1)$-resolving & 10 & $1.10\ee{7}$ & 13.1 GiB & 29.7 GiB \\
        $\SO(5)$ DQCP & $\SO(3)$-resolving & 10 & $1.60\ee{5}$ & 130 MiB & 127 MiB \\
        $\SO(5)$ DQCP & $\SO(3)$-resolving & 12 & $1.08\ee{7}$ & 12.7 GiB & 12.0 GiB \\
        \hline
        $\Sp(3)$, filling-2 & $\rU(1)$-resolving & 7 & $2.45\ee{6}$ & 2.66 GiB & 4.52 GiB \\
        $\Sp(3)$, filling-2 & $\SO(3)$-resolving & 7 & $5.81\ee{4}$ & 738 KiB & 291 MiB \\
        $\Sp(3)$, filling-2 & $\SO(3)$-resolving & 9 & $1.48\ee{7}$ & 17.8 MiB & 15.4 GiB \\
        \hline
        $\SU(2)_1$-Higgs & $\SO(3)$ full & 8 & $3.97\ee{3}$ & 1.28 GiB & 1.46 GiB \\
        $\SU(2)_1$-Higgs & $\rU(1)$-resolving & 9 & $9.05\ee{6}$ & 1.94 GiB & 15.5 GiB \\
        $\SU(2)_1$-Higgs & $\SO(3)$ quasi-hole & 9 & $4.12\ee{3}$ & 32.7 MiB & 15.5 MiB \\
        $\SU(2)_1$-Higgs & $\SO(3)$ quasi-hole & 12 & $7.72\ee{5}$ & 17.8 GiB & 12.0 GiB \\
        \hline
        Majorana fermion & $\SO(3)$ full & 12 & $4.08\ee{3}$ & 4.61 GiB & 8.61 GiB\\
        Majorana fermion & $\rU(1)$-resolving & 13 & $3.23\ee{6}$ & 360 MiB & 5.71 GiB\\
        Majorana fermion & $\SO(3)$ quasi-hole & 13 & $2.84\ee{3}$ & 31.7 MiB & 9.48 MiB\\
        Majorana fermion & $\SO(3)$ quasi-hole & 16 & $9.01\ee{4}$ & 4.14 GiB & 583 MiB \\
        \hline\hline
    \end{tabular}
    \caption{Statistics on the Hilbert-space dimension and the memory usage for the $\SO(5)$ deconfined criticality, the $\Sp(N)$-symmetric CFTs, $\SU(2)_1$-Higgs theory, and the free Majorana fermion using the $\rU(1)$-resolving and the $\SO(3)$-resolving ED. For the $\SO(3)$-resolving ED, we consider the sector of the $\SO(3)$ spin-0, flavour-singlet sector. The discrete $\BZ_2$ symmetries are taken into account for the $\rU(1)$-resolved Hilbert-space dimension and not taken into account for the $\SO(3)$-resolved dimension. For FQH transitions, both the full space and the quasi-hole space are listed. }
    \label{tbl:bench_other}
\end{table}

\section{Identifying Heavy Primaries of Ising CFT}
\label{sec:results_prim}

The $\SO(3)$-resolving ED enables us to calculate higher spectrum with a specific spin. As a result, we have identified 95 primary operators of the 3D Ising CFT with Lorentz spin $l\leq 2$. We calculate the 2000 lowest-lying eigenstates for the $(l,P,Z)=(0,+,\pm)$ sectors\footnote{To calculate a large number of eigenstates in Lanczos, we adopt a locked Lanczos algorithm that removes converged Ritz vectors from the active Krylov subspace, retaining them only for orthogonalization against newly generated vectors. The code is publicly available at the forked GitHub repository \href{https://github.com/FuzzifiED/KrylovKit.jl}{\texttt{FuzzifiED/KrylovKit.jl}}.} and 300 lowest-lying eigenstates for the $(l,P,Z)=(0,-,\pm),(1,\pm,\pm),(2,\pm,\pm)$ sectors up to system size $N_m=20$. These cover the states up to $\Delta_{\max}\approx 22$ (parity-even) or $19$ (parity-odd) for $l=0$, $\Delta_{\max}\approx 16.5$ for $l=1$, and $\Delta_{\max}\approx 15$ for $l=2$. We also calculate the lowest 50 states for each $l=0$ sector for $N_m=22$.

Because of irrelevant perturbations, each Hamiltonian eigenstate can be a linear combination of different primaries and descendants. We isolate the primary states using the conformal generators~\cite{Fardelli2024,Fardelli2026}. In the infra-red, the first moment of the Hamiltonian density $H(\br)$ is identified with the generator
\begin{equation}
    \Lambda^\mu=P^\mu+K^\mu=\int\rd^2\br\,r^\mu H(\br;U'_0,U'_1,h').
\end{equation}
The parameters $U'_0,U'_1,h'$ differ from their conformal-point values to account for total-derivative terms. We determine them by minimizing the norm of $\Lambda^\mu\|0\rangle$ while imposing the normalization condition $\langle\partial\epsilon\|\Lambda\|\epsilon\rangle=\sqrt{2\Delta_\epsilon}$. The special conformal transformation (SCT) generator is then obtained from its commutator with the dilatation operator:
\begin{equation*}
    K^\mu=\frac{1}{2}\left(\Lambda^\mu-[D,\Lambda^\mu]\right)
\end{equation*}
where the dilatation operator is obtained by calibrating the Hamiltonian against the $\sigma$ state: $D=(H-E_0)/(v/R)$, with $v/R=(E_\sigma-E_0)/\Delta_\sigma$. The primaries should have $|K^\mu|\Phi\rangle|^2=\langle\Phi|\Xi|\Phi\rangle=0$ where $\Xi=|K|^2=P_\mu K^\mu$.

We filter the primary operator by choosing the lowest eigenvalues of $\Xi=|K^\mu|^2$ within the low-lying energy eigen-space. This method was first developed in Ref.~\cite{Fardelli2026}. To identify the primaries in each $(l,P,Z)$ sector, we diagonalize $\Xi$ within the space spanned by the Hamiltonian eigenstates $|E_i\rangle$.
\newcommand{\bU}{\mathbf{U}}
\newcommand{\bT}{\mathbf{T}}
\newcommand{\bD}{\mathbf{D}}
\begin{align}
    \Xi_{ij}&=\langle E_i|\Xi| E_j\rangle,&|\xi_\alpha\rangle&=\bU_{\alpha i}|E_i\rangle,&\Xi|\xi_\alpha\rangle&=\xi_\alpha|\xi_\alpha\rangle
\end{align}
where $\xi_\alpha$ is the $\alpha$-th smallest eigenvalue of $\Xi=|K|^2$.  
The spectrum should exhibit a separation: the first $N_\text{prim}$ eigenvalues are close to zero, whereas the rest remain finite. The corresponding $N_\text{prim}$ eigenvectors therefore span the candidate primary subspace. Finally, we diagonalize the dilatation operator within this subspace
\begin{align}
    &\bD_{\alpha\beta}=\langle \xi_\alpha|D|\xi_\beta\rangle=\bU_{\alpha i}(\tilde{\Delta}_i\delta_{ij})\bU^\dagger_{j\beta}\qquad(\alpha,\beta\leq N_\text{prim},\quad\tilde{\Delta}_i=\tfrac{R}{v}(E_i-E_0))\nonumber\\
    &\bT_{n\alpha}\bD_{\alpha\beta}\bT^{\dagger}_{\beta m}=\Delta_n\delta_{nm}\nonumber\\
    &|\Phi_n\rangle=\bT_{n\alpha}|\xi_\alpha\rangle=\bT_{n\alpha}\bU_{\alpha i}|E_i\rangle
\end{align}
The resulting eigenstates $|\Phi_n\rangle$ are identified as the primary states. Their scaling dimensions are the eigenvalues $\Delta_n$. 

\begin{figure}[htbp]
    \centering
    \includegraphics[width=0.48\linewidth]{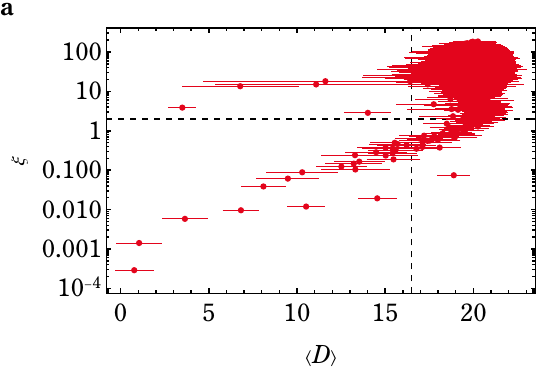}
    \includegraphics[width=0.48\linewidth]{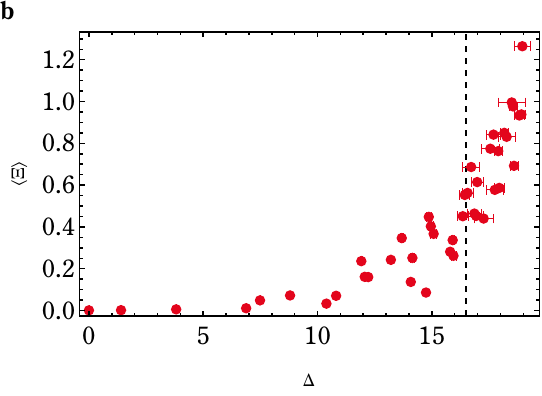}
    \caption{(a) Eigenvalues $\xi_\alpha$ of $\Xi=|K|^2$ in the $(l,P,Z)=(0,+,+)$ sector at $N_m=20$, plotted against the expectation value of the dilatation operator $\langle D\rangle=\langle\xi_\alpha|D|\xi_\alpha\rangle$. The error bars denote the standard deviation of the dilatation operator $\sqrt{\langle D^2\rangle-\langle D\rangle^2}$. The grid lines denote the cut-off value for $\xi$ and scaling dimension $\Delta$. (b) Scaling dimensions $\Delta$ and expectation values $\langle \Xi\rangle$ of the identified primaries. The error bars show ten times the standard deviations of $\Delta$ upon varying cut-offs $\xi_c$ and $\Delta_{\max}$. Results for other sectors are plotted in Figures~\ref{fig:k2eig} and \ref{fig:deig}.}
    \label{fig:eig_k2}
\end{figure}

As an example, we consider the $(l,P,Z)=(0,+,+)$ sector at $N_m=20$. The 2000 eigenstates within the sector cover up to $\Delta_{\max}=22$. We examine the eigenstates of $\Xi$. For each eigenstate, we study its profile by measuring the expectation value $\langle D\rangle=\langle\xi_\alpha|D|\xi_\alpha\rangle$ and the standard deviation $\sigma_D=\sqrt{\langle D^2\rangle-\langle D\rangle^2}$ of the dilatation operator $D$~(Figure~\ref{fig:eig_k2}a). For $\langle D\rangle$ smaller than a cut-off $\Delta_c\approx 16.5$, we observe a clear separation between primaries and descendants at $\xi_c=2$. For $\Delta>\Delta_c$ the gap between primaries and descendants gradually closes. The resulting mixing is likely a finite-window effect associated with the proximity to $\Delta_{\max}$, where the computed states can hybridize with higher-lying states beyond the accessible spectrum.

We therefore identify the space spanned by all the $\xi<\xi_c$ states as the candidate primary subspace. In this space, the $\Xi$-eigenstates mix conformal operators within a relatively small range of $\Delta$, with $\sigma_D\approx 1$. Diagonalizing $D$ within this space gives all the primary states and their scaling dimensions. We test the stability of their $\Delta$ by varying the cut-off $\xi_c$ and lowering the maximal scaling dimension $\Delta_{\max}$. The error bars in Figure~\ref{fig:eig_k2}b show ten times the standard deviation of $D$ for each primary upon varying $\xi_c$ from $2$ to $4/3$ and $3$, and upon retaining the states only to $\Delta'_{\max}=20$. For primaries with $\Delta<\Delta_c=16.5$, the variations of $\Delta$ upon changing the truncation scheme are within 0.02, and their $\langle\Xi\rangle<0.6$; the variation becomes much larger and $\langle\Xi\rangle$ becomes higher for $\Delta>\Delta_c$. We therefore conclude that the primaries below $\Delta_c\approx 16.5$ can be reliably determined. 

\begin{figure}[htbp]
    \centering
    \includegraphics[width=0.24\linewidth]{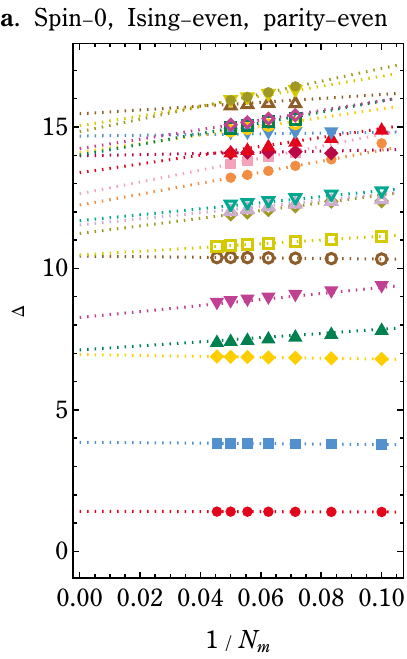}
    \includegraphics[width=0.24\linewidth]{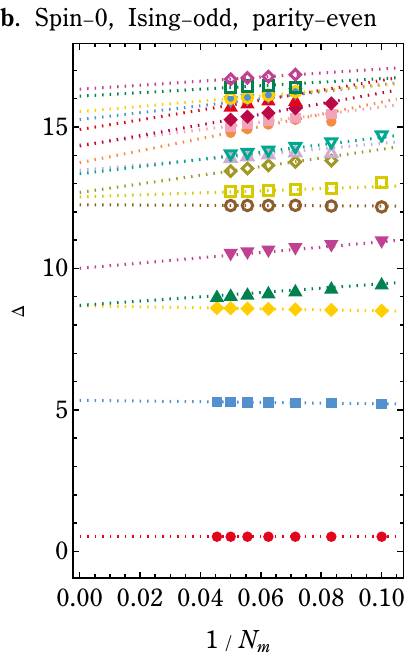}
    \includegraphics[width=0.24\linewidth]{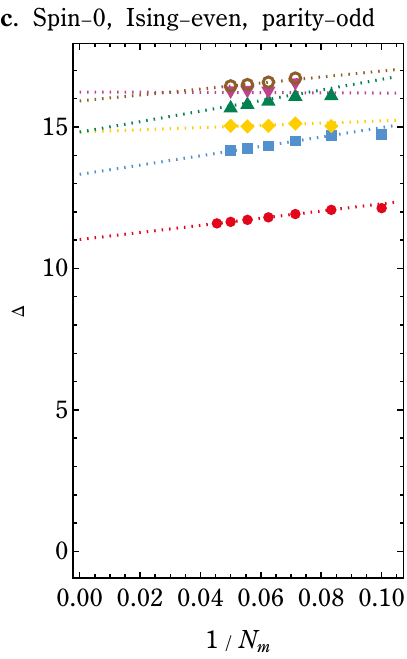}
    \includegraphics[width=0.24\linewidth]{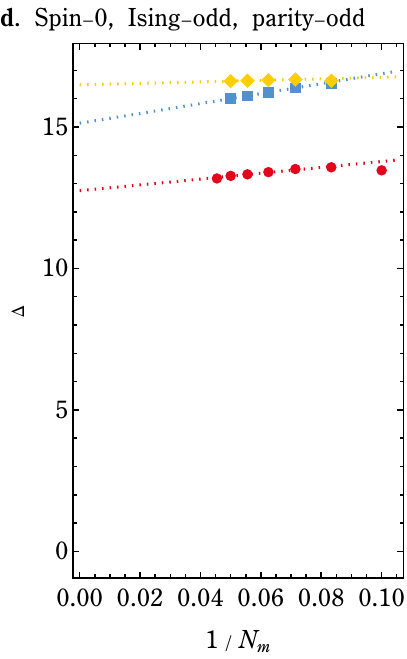}\\
    \includegraphics[width=0.24\linewidth]{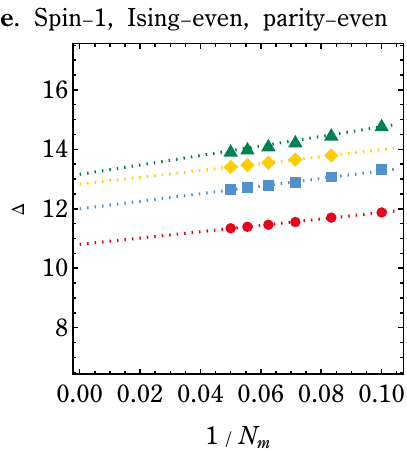}
    \includegraphics[width=0.24\linewidth]{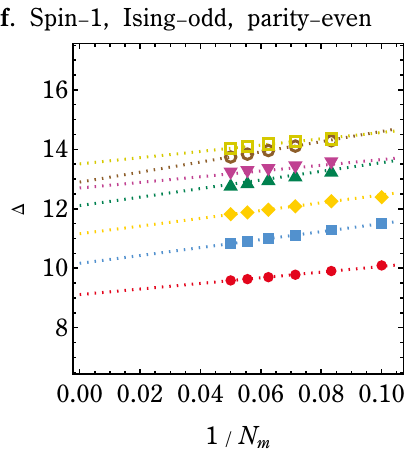}
    \includegraphics[width=0.24\linewidth]{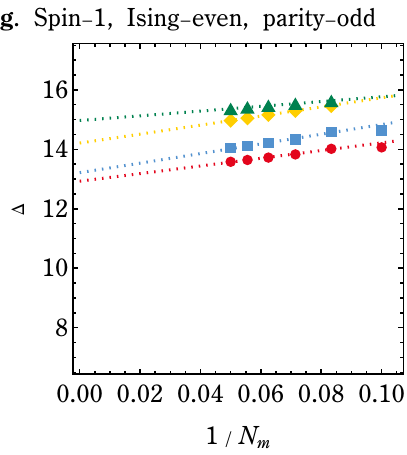}
    \includegraphics[width=0.24\linewidth]{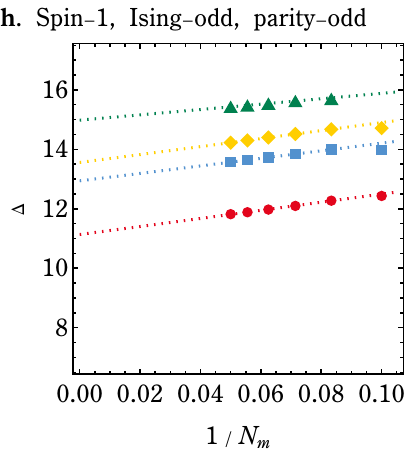}\\
    \includegraphics[width=0.24\linewidth]{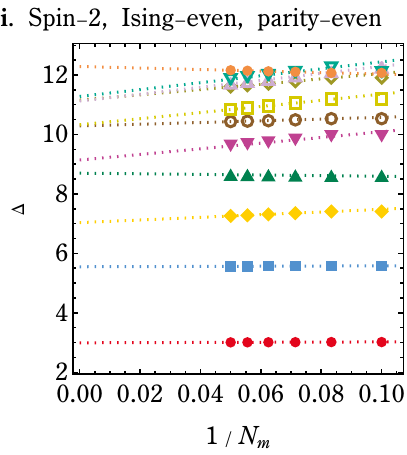}
    \includegraphics[width=0.24\linewidth]{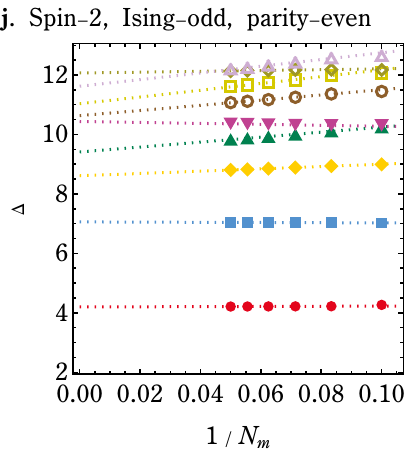}
    \includegraphics[width=0.24\linewidth]{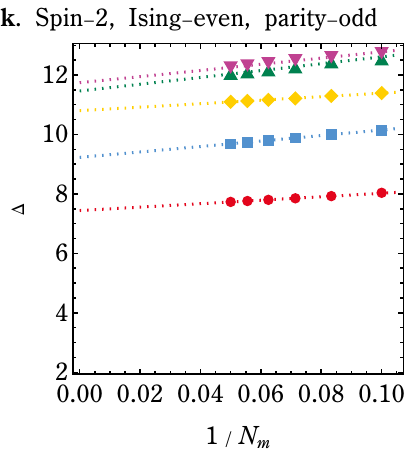}
    \includegraphics[width=0.24\linewidth]{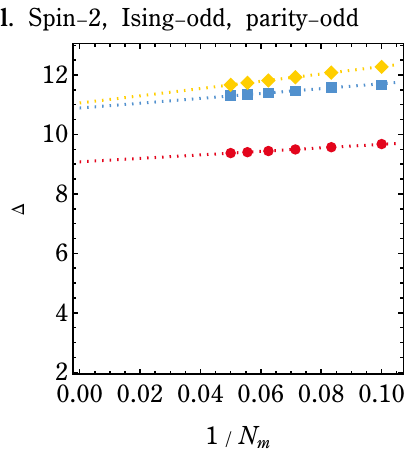}
    \caption{The finite-size extrapolation of scaling dimensions of conformal primaries with spin $l=0,1,2$ and various Ising and space-time-parity sectors, retaining only the $N^{-1.01}$ contribution of $C_{\mu\nu\rho\sigma}$.}
    \label{fig:prim_dim}
\end{figure}

Using this method, we have identified 48 spin-0 primaries up to $\Delta\approx 16.5$\footnote{This bound refers to the finite-size values. A primary with $\Delta>16.5$ at finite size may still have $\Delta<16.5$ in the thermodynamic limit.} including 9 parity-odd ones, 19 spin-1 primaries up to $\Delta\approx 14$ (parity-even) or $15.5$ (parity-odd), and 28 spin-2 primaries up to $\Delta\approx 12.5$. To extract their scaling dimensions in the thermodynamic limit, we perform a finite-size extrapolation. As the deformations from $\epsilon$ and $\epsilon'$ have been tuned away~\cite{Fardelli2024}, the leading contributions to the finite-size effect are sub-leading operators $C_{\mu\nu\rho\sigma}$ and $T'_{\mu\nu}$.\footnote{The coupling to curvature has the power $N^{-1}$ very close to the contribution of $C_{(4)}$. We therefore combine their contributions together and do not make separate fitting.} The fitting ansatz is 
\begin{align}
    \Delta(N_m)&=\Delta+\alpha_{C_{(4)}} N_m^{-(\Delta_{C_{(4)}}-3)/2}+\alpha_{T'} N_m^{-(\Delta_{T'}-3)/2}+\cdots\nonumber\\
    &=\Delta+\alpha_{C_{(4)}}N_m^{-1.01}+\alpha_{T'}N_m^{-1.26}+\cdots
\end{align}
We consider three estimates 
\begin{enumerate}[nosep]
    \item extrapolating including only $C_{\mu\nu\rho\sigma}$,
    \item extrapolating including both $C_{\mu\nu\rho\sigma}$ and $T'_{\mu\nu}$, and
    \item directly taking the value at the largest system size.
\end{enumerate}
We report the value including only $C_{\mu\nu\rho\sigma}$ and take the root-mean-square of the differences with the other two estimates as the error bar. We plot the finite-size extrapolations in Figure~\ref{fig:prim_dim} and list the estimates in Table~\ref{tbl:prim_dim}. Our results show good consistency with various existing results obtained by other methods, \emph{viz.}~conformal bootstrap~\cite{Poland2018Bootstrap,Rychkov2023Bootstrap,Simmons-Duffin:2016wlq,Chang:2024whx,Henriksson:2022gpa} and $\epsilon$-expansion~\cite{Antipin:2025ilv,Henriksson:2022gpa,Henriksson:2025vyi}.\footnote{For perturbative results known through orders $\epsilon^2$, $\epsilon^3$, and $\epsilon^5$, the Pad\'e approximants $[1,1]$, $[1,2]$, and $[2,3]$ are used. An Ising-even spin-2 operator of from $\partial^2\Box^2\phi^4$ with scaling dimension 8.11 at order $\epsilon^1$ is missing from our spectrum. We thank Johan Henriksson for providing guidance for these results.} For the primaries previously discovered on the fuzzy sphere (\textit{e.~g.}~the spin-2 Ising-even parity-odd primary at $7.41(18)$ reported in Ref.~\cite{Fardelli2026}), our results show agreement. Many other primaries are observed for the first time.

We have made one of the first reports on the spin-0 parity-odd primaries. The first Ising-even parity-odd primary $\epsilon_-$ is the second lowest operator of the same symmetry sector. Our result $\Delta_{\epsilon_-}=11.0(4)$ agrees with the bound $10.88<\Delta_{\epsilon_-}\CB<10.93$ from the stress-tensor conformal bootstrap~\cite{Chang:2024whx} under the assumption that the second such primary has $\Delta_{\epsilon'_-}>11.5$, and our observation $\Delta_{\epsilon'_-}=13.3(6)$ supports this assumption. We remark that the original paper on fuzzy-sphere~\cite{Zhu2022} reported the first parity-odd scalars in both the Ising even and odd sectors as primaries, but we find that both of them are indeed descendants, as they have large $\langle\Xi\rangle=\langle|K|^2\rangle$ values. For example, the first scalar in the parity-odd, Ising-even sector has $\Delta\approx 10$, plausibly the descendant $\partial_\mu\partial_\nu T_-^{\mu\nu}$ with $\Delta_{T_-}=7.4(2)$.

We make one of the first reports on the spin-1 primaries. Our observation for the first Ising-odd parity-even primary $\Delta_{\sigma_\mu}=9.1(4)$ agrees with the observation $\Delta_{\sigma_\mu}\approx 9.4$ by conformal bootstrap in the $\sigma\times\epsilon$ channel~\cite{Henriksson:2022gpa}. We also find the first Ising-even parity-even spin-1 primary $\Delta_{V_\mu}=10.8(5)$; this operator is related to the structural explanation of the conformal symmetry of the 3D Ising transition~\cite{Meneses:2018xpu}. It would be interesting to compare our result with the bootstrap in the OPE channel of two different primaries with the same Ising parity, such as $\epsilon\times\epsilon'$. 

Several of the parity-even scalars can be identified within the $\phi^n$ series. In addition to the operators with $n\leq6$ reported by conformal bootstrap~\cite{Simmons-Duffin:2016wlq}, we observe the $n=7,8,9,10$ operators with scaling dimensions $8.68(6),10.43(9),12.53(15)$, and $14.68(6)$, respectively. These values are comparable with the five-loop $\epsilon$-expansion results~\cite{Antipin:2025ilv,Sannino:2026ena}, although the fuzzy-sphere estimates are generally slightly smaller. Curiously, these operators tend to exhibit smaller finite-size corrections and smaller values of $\langle\Xi\rangle$ than other operators at similar scaling dimensions, and form a clearly separated branch among the $\Xi$ eigenstates (Fig.~\ref{fig:eig_k2}a). In \S~\ref{sec:results_apd}, we investigate the structure of this branch in more detail.

\section{Semi-Classical $\phi^n$ Branch and Area-Preserving Diffeomorphism on Sphere}

\label{sec:results_apd}

We now turn to a special structure associated with the $\phi^n$ family. In the Ising CFT, the operators $\BI,\sigma,\epsilon,\square\sigma,\epsilon',\sigma',\ldots$ are naturally associated with the operators $\phi^n$.\footnote{The operator $\phi^3$ being the descendant of $\phi$ ($\sigma$) in the Ising CFT is a consequence of the equation of motion $\square\phi=\frac{1}{6}\lambda\phi^3$ of the $\phi^4$ QFT $\mathcal{L}=\frac{1}{2}(\partial_\mu\phi)^2+\frac{\lambda}{4!}\phi^4$.} Our observation is that, in the fuzzy-sphere Hilbert space, these states appear to form a distinguished branch with a special wave-function structure. As we show below, this structure is closely tied to area-preserving diffeomorphisms (APD) of the sphere and admits a rather semiclassical picture.

Even though the 3D Ising CFT is a bosonic QFT, the fuzzy-sphere model realizing it is constructed entirely from fermionic degrees of freedom, which remain gapped at the CFT point. This is reminiscent of the emergence of the Heisenberg spin model from the fermionic Hubbard model~\cite{Anderson:1959} and, more generally, of the emergence of spin models in electronic systems. It is therefore natural to ask whether there exists a Mott-insulating limit in which the fermion density is spatially uniform and its fluctuations are completely frozen,
\begin{equation}
    \langle (n^0(\br))^2 \rangle-\langle n^0(\br) \rangle^2=0,\qquad\forall \br\in S^2,
\end{equation}
such that an effective spin model for the fuzzy-sphere Ising CFT can be derived.

For the fuzzy sphere Ising model with $N_m$ fermions, there are only $N_m+1$ uniform states with zero charge fluctuations, in contrast to the exponentially large number of such states in a lattice model. They are the zero-energy states of the term $\int_{S^2} \rd^2\br\,(n^0(\br)-1/\sqrt{4\pi})^2$. These states form a spin-$N_m/2$ multiplet under the $\SU(2)_f$ flavour symmetry where the two fermion flavours transform in the fundamental representation. They can be written compactly as
\begin{equation}
|D_k\rangle=\frac{1}{k!}\binom{N_m}{k}^{-1/2}\left(S^+\right)^k\left(\prod_m c^\dagger_{m\downarrow}|0\rangle\right),
\end{equation}
where
\begin{equation}
    S^+=\sum_mc^\dagger_{m\uparrow}c_{m\downarrow}
\end{equation}
is the raising operator of the $\SU(2)_f$. It is proportional to the density operator $n^+_{l=0,m=0}$. Thus, $|D_0\rangle$ is a trivially polarized state, while $|D_k\rangle$ is obtained from it by flipping $k$ spins. Moreover, the states $|D_k\rangle$ are all $\SO(3)$ scalars, and they are the only states on the fuzzy sphere that are invariant under the APD of $S^2$. Consequently, their (spin) correlation functions are trivial, \emph{i.~e.}~constants independent of the distance. Therefore, there is no Mott-insulating limit for the physics of 3D Ising CFT on the fuzzy sphere. 

\begin{figure}
    \centering
    \includegraphics[width=0.8\linewidth]{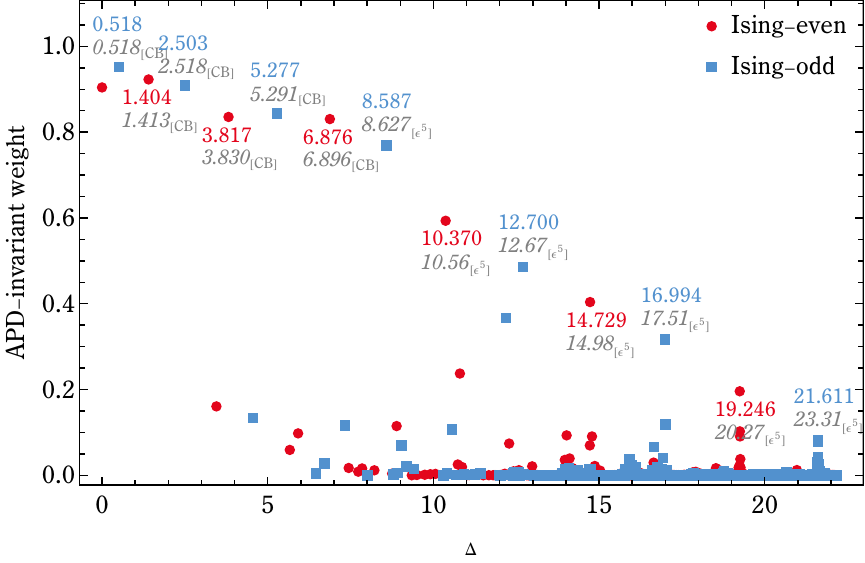}
    \caption{The wave-function weights on the APD-invariant sector. Here we show the 2000 lowest eigenstates of the Ising CFT Hamiltonian in the sectors $(l,P,Z)=(0,+,+)$ and $(l,P,Z)=(0,+,-)$ at $N_m=20$, respectively.  The scaling dimensions in the bracket are from the numerical bootstrap~\cite{Poland2018Bootstrap} or $\epsilon$-expansion~\cite{Antipin:2025ilv}.}
    \label{fig:APDweights}
\end{figure}

Nevertheless, these APD-invariant states $|D_k\rangle$ have direct implications for the fuzzy sphere Ising CFT, particularly for the states in the $\phi^n$ branch, including the ground state. One way to see this is to compute, for each Hamiltonian eigen state $|\psi\rangle$, the total weight of the wave function in the subspace spanned by the $|D_k\rangle$ states,
\begin{equation}
w_\text{APD}=\sum_{k=0}^{N_m} \left|\langle D_k|\psi\rangle\right|^2.
\end{equation}
Figure~\ref{fig:APDweights} shows the APD-invariant weights $w_\text{APD}$ of the low-lying states of the $N_m=20$ model. Since the APD-invariant subspace is only $N_m+1=21$-dimensional, whereas the $\SO(3)$ scalar sector has a dimension of roughly four million, one would expect a typical state to have a negligibly small $w_\text{APD}$, as is indeed evident in the figure. Nevertheless, the $\phi^n$ states have remarkably large weights\footnote{The weights decrease slowly as the system size $N_m$ increases, \textit{e.~g.}~see Figure~\ref{fig:WF_weights} in the Appendix.} in the APD-invariant subspace. In particular, the low-lying states ($n=0,1,2,3$) have $w_\text{APD}\gtrsim 0.9$. This is rather surprising, not only because the APD-invariant subspace constitutes a negligible fraction of the exponentially large Hilbert space, but also because the APD-invariant states themselves have trivial correlation functions. For larger $n$, around $n\gtrsim 9$, $w_\text{APD}$ begins to decrease as the $\phi^n$ states increasingly mix with other CFT states. This mixing reduces the APD weight of the main bare $\phi^n$ states and redistributes it into a tail of other CFT states with smaller but non-zero $w_\text{APD}$. Despite this mixing, the $\phi^n$ states can still be readily identified up to $n=13$, and their bare scaling dimensions are in good agreement with the $\epsilon$-expansion and bootstrap results, where available. It would be interesting to develop a systematic way to disentangle the mixing between the higher-$n$ $\phi^n$ states and nearby CFT states.

\begin{figure}
    \centering
    \includegraphics[width=0.48\linewidth]{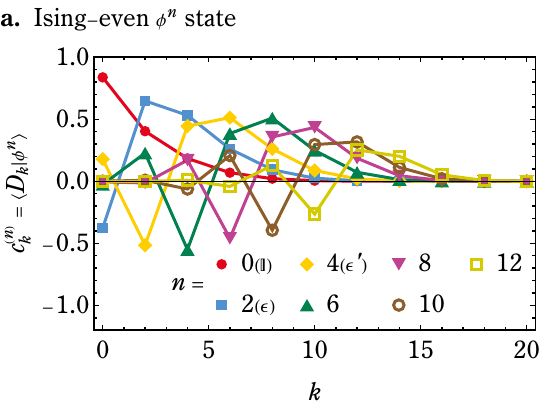}
    \includegraphics[width=0.48\linewidth]{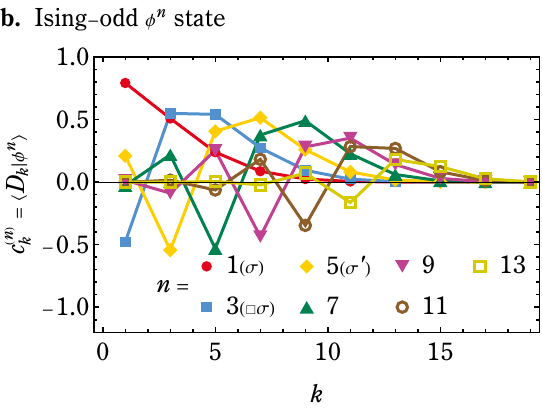}
    \caption{Wave-function amplitudes $c_k^{(n)}=\langle D_k|\phi^n\rangle$ of $|\phi^n\rangle$ states in the APD-invariant subspace $|D_k\rangle$ measured at $N_m=20$.}
    \label{fig:phin_distribution}
\end{figure}

Since each $|D_k\rangle$ is obtained by flipping $k$ spins relative to the fully polarized state $|D_0\rangle$, it is interesting to study the wave-function amplitudes $c_k^{(n)}=\langle D_k|\phi^n\rangle$, as shown in Figure~\ref{fig:phin_distribution}. The $\BI$ and $\phi\sim\sigma$ are clearly dominated by the fully polarized state and the one-spin-flip state, respectively, with their amplitudes spreading to higher-spin-flip configurations in a rapidly decaying manner. For higher $\phi^n$ states, the structure becomes more intricate: the dominant support shifts progressively toward larger $k$, although it is not centered exactly at $k=n$ (except for $n=2$), and is accompanied by a decaying tail toward still larger $k$. The amplitudes also exhibit sign oscillations at smaller $k$, which can be understood as reflecting the tendency of different $\phi^n$ states to remain mutually orthogonal within the APD-invariant subspace. These observations suggest an underlying semi-classical structure of the $\phi^n$ states in the fuzzy-sphere model, whose implications would be interesting to explore in future work.

\section{Quasi-Hole-Space Projection for Fractional Quantum Hall Transitions}
\label{sec:results_trunc}

\newcommand{\inter}{\text{inter}}
\newcommand{\intra}{\text{intra}}
\newcommand{\Ising}{\text{Ising}}
\newcommand{\FQH}{\text{FQH}}
\newcommand{\qh}{\text{qh}}

In this section, we turn to fuzzy-sphere CFTs involving fractional quantum Hall states. The literature has so far explored two distinct settings. The first concerns phase transitions between a fractional quantum Hall state and another phase, which may be another fractional or integer quantum Hall state or a symmetry-broken state. Such transitions are typically described by Chern-Simons-matter CFTs, and a prototypical example has recently been studied on the fuzzy sphere~\cite{Zhou2025Jul}. The second uses fractional quantum Hall states as a regulator, or background, to realize symmetry-breaking transitions such as the Ising transition~\cite{Voinea2024}, in analogy with the original construction of the Ising transition on top of an integer quantum Hall background~\cite{Zhu2022}.

This class of fuzzy-sphere CFTs opens up a rich territory for exploration. Despite the recent progress, their numerical study faces several challenges. First, for the same number of particles, fractional quantum Hall states typically have a much larger Hilbert space than integer quantum Hall states, restricting accessible calculations to considerably smaller system sizes. Second, the energy spectra of these fuzzy-sphere CFTs often contain magneto-roton excitations of the underlying fractional quantum Hall state, which are not part of the CFT spectrum and may contaminate the identification of CFT states~\cite{Voinea2024,Zhou2025Jul}.\footnote{These magneto-roton excitations are gapped, but they can have comparable gaps with CFT states at small sizes because the latter have $\Delta/R$ gaps.} Below, we show that these challenges can be addressed by a physically motivated projection onto the quasi-hole zero-mode spaces associated with model fractional quantum Hall wave functions and their exact parent Hamiltonians, including the Laughlin, Halperin, and Moore-Read states~\cite{Laughlin1983Anomalous,Halperin:1983zz,Moore1991Pfaffian,Haldane1983FQHE,Bernevig:2008rda}.

\subsection{Chern-Simons-Matter CFT}

We first analyse the phase transition between the $\nu=1/2$ bosonic Laughlin state and a trivially gapped phase~\cite{Zhou2025Jul}. This transition can be described by several (conjecturally) dual Chern-Simons-matter theories, including a critical scalar field coupled to an $\SU(2)_1$ Chern-Simons gauge field. We therefore refer to it as the $\SU(2)_1$ Higgs transition. The fuzzy-sphere model of this transition consists of two charge-1 fermionic flavours $f^\sigma$ and a charge-2 bosonic flavour $b$, and the model Hamiltonian is \begin{equation}
    H=\int\rd^2\br\left[U_bn_b^2+U_{bf}n_bn_f+U_fn_f^2-t(b^\dagger f^1f^2+\text{h.~c.})+\mu n_f\right],
    \label{eq:hmt}
\end{equation}
where $n_f=f_i^\dagger f^i$ and $n_b=b^\dagger b$. We take the same parameters $U_b=4,U_{bf}=4,U_f=1,t=1/2$ as in Ref.~\cite{Zhou2025Jul}. The number of fermionic orbitals is $N_{mf}=2s+1$, while the number of bosonic orbitals is $N_{mb}=4s+1$, and the total electric charge $Q=N_f+2N_b=4s+2$ is conserved.  The transition is driven by tuning $\mu$: (1) large positive $\mu$ suppresses the fermionic occupation and favours the bosonic sector (\textit{e.~g.}~$N_f=0$, $N_b=2s+1$), realizing a bosonic Laughlin state at filling $\nu_b=1/2$; (2) large negative $\mu$ favours the fermionic sector (\textit{e.~g.}~$N_f=4s+2$, $N_b=0$), realizing a fermionic integer quantum Hall state at filling $\nu_f=2$; and (3) the critical point was found at $\mu_c=0.312$.

The entire Hilbert space of the model can be written as a direct sum of tensor products of bosonic and fermionic sectors,
\begin{equation}
\cH(s)=\bigoplus_{N_b=0}^{2s+1}\cH_b(N_b)\otimes \cH_f(N_f=4s+2-2N_b).
\end{equation}
Since $N_b$ and $N_f$ are not separately conserved, a generic state can have support across different $N_b$ sectors in this decomposition. In the limit $\mu\to\infty$, the system is confined to the purely bosonic sector $\cH_b(N_b=2s+1)$, and the Hamiltonian reduces to a single interaction term, $n_b^2$. This is the famous Haldane-pseudopotential parent Hamiltonian for the $\nu_b=1/2$ bosonic Laughlin wave function~\cite{Haldane1983FQHE}. More precisely, the normal-ordered Hamiltonian $H_\FQH=\int\rd^2\br\,:n_b^2:$ takes the Haldane-pseudopotential form and is positive semidefinite, with the Laughlin wave function as its unique zero mode in $\cH_b(N_b=2s+1)$. 

As $\mu$ is lowered toward the critical value $\mu_c=0.312$, the eigenstates acquire quantum fluctuations into the fermionic sectors and become boson-fermion mixtures, with their wave functions developing finite weight in sectors with $N_b<2s+1$, namely $\cH_b(N_b)\otimes\cH_f(N_f=4s+2-2N_b)$. From the perspective of the bosonic component, these $N_b<2s+1$ sectors can be viewed as hole-doped sectors relative to the Laughlin wave function. It is well known that, in these sectors, the pseudopotential Hamiltonian $H_\FQH$ has many additional exact zero modes, described by Laughlin quasi-hole wave functions~\cite{Haldane1983FQHE}. These zero-energy quasi-hole states have a simple physical interpretation: a bosonic hole fractionalizes into two Laughlin quasi-holes, each an anyonic excitation carrying fractional charge $1/2$ relative to that of the original bosons~\cite{Laughlin1983Anomalous}. Although the zero-mode quasi-hole space $\cH_b^\qh(N_b)$ occupies only a small fraction of the full bosonic sector $\cH_b(N_b)$, it includes all the anyonic excitations of the Laughlin state. One may conjecture that $\cH_b^\qh(N_b)$ may continue to play an important role as $\mu$ decreases, including at the critical point. 

To quantify the importance of the quasi-hole space $\cH_b^\qh(N_b)$ at the $\SU(2)_1$ Higgs transition, we measure the wave function weights of low-energy states at the transition in the truncated space 
\begin{equation}
    \cH^\qh(s)=\bigoplus_{N_b=0}^{2s+1}\cH^\qh_b(N_b)\otimes \cH_f(N_f=4s+2-2N_b).   
\end{equation}
Across different system sizes, the CFT states, such as $\BI,S,\partial^\mu S,J^\mu$, have weights in the quasi-hole space very close to unity, exceeding 0.9996 (Table~\ref{tbl:trunc_wt_su2}). By contrast, non-CFT states such as the magneto-roton state (\emph{e.~g.}~the lowest spin-4 flavour-singlet state) have substantially smaller weights in the quasi-hole space. This weight increases with system size as the magneto-roton state increasingly mixes with CFT states in the same sector when its rescaled scaling dimension moves upward. These results strongly suggest that $\cH^\qh(s)$ provides a natural restricted Hilbert space for studying the $\SU(2)_1$ Higgs transition. 

\begin{table}[htbp]
    \centering
    \renewcommand{\tabcolsep}{10pt}
    \begin{tabular}{r|cccc}
        \hline\hline
        $N_{mf}$ & 8 & 7 & 6 & 5 \\ 
        \hline
        $\BI$            & 0.99998 & 0.99998 & 0.99999 & 0.99999 \\
        $S$              & 0.99993 & 0.99994 & 0.99995 & 0.99997 \\
        $\partial^\mu S$ & 0.99983 & 0.99985 & 0.99987 & 0.99990 \\
        $J^\mu$          & 0.99969 & 0.99970 & 0.99972 & 0.99976 \\
        roton            & 0.36372 & 0.30596 & 0.26384 & 0.23386 \\
        \hline\hline
    \end{tabular}
    \caption{The weight of different states within the quasi-hole space $\cH^\qh$ for the $\SU(2)_1$-Higgs transition, measured by the overlap $\langle\Phi|\Pi_\qh|\Phi\rangle$ for different system sizes. $\Pi_\qh$ is the projector for the quasi-hole space. ``Roton'' denotes the lowest $l=4,s=0$ state. The model and parameters are taken from Ref.~\cite{Zhou2025Jul}.}
    \label{tbl:trunc_wt_su2}
\end{table}

\begin{figure}[htbp]
    \centering
    \includegraphics[width=0.48\linewidth]{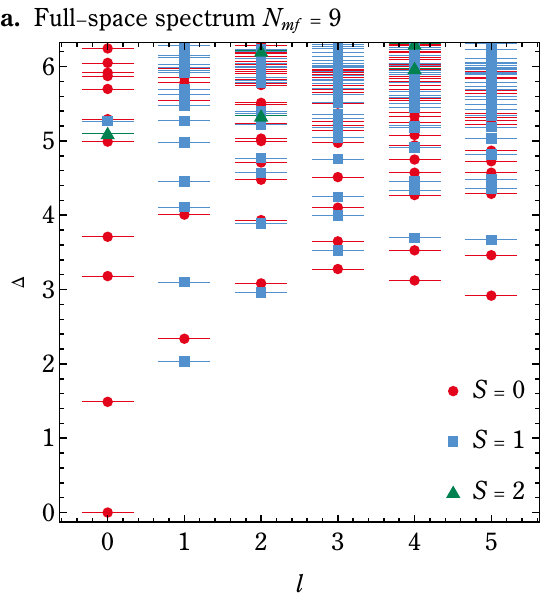}
    \includegraphics[width=0.48\linewidth]{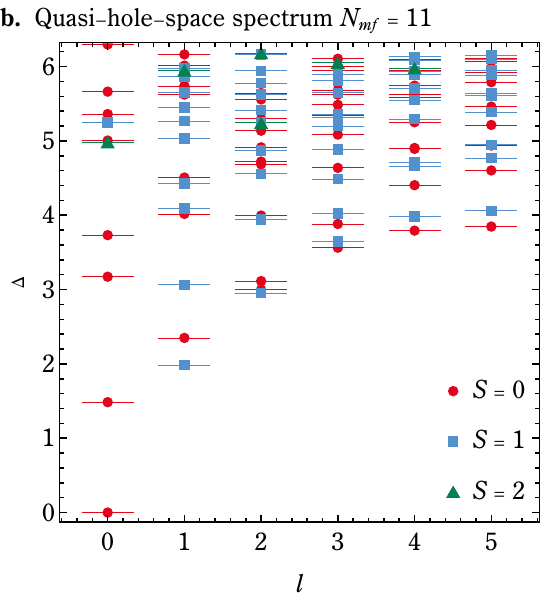}
    \caption{The spectrum of the $\SU(2)_1$-Higgs transition within (a) full Hilbert space at $N_m=9$ and (b) projected quasi-hole space at $N_m=11$. Different markers denote different flavour-$\SO(3)$-spin $S$. The model and parameters are taken from Ref.~\cite{Zhou2025Jul}. }
    \label{fig:trunc_spec_su2}
\end{figure}

Therefore, we directly study the $\SU(2)_1$ Higgs transition by projecting onto the quasi-hole space $\cH^\qh(s)$. In practice, this Hilbert-space projection is implemented by constructing the bosonic sector of the Hilbert space from Jack states squeezed from the $(1,2)$-admissible root configurations~\cite{Bernevig:2007nek,Bernevig:2008rda,Bernevig2008,Bernevig2009} and combining it with the fermionic sector. Figure~\ref{fig:trunc_spec_su2} shows the energy spectrum computed in the projected quasi-hole space. The scaling dimensions of the low-lying CFT states are almost unchanged from their values before projection, confirming that the projection has little effect on the CFT states. On the other hand, the projection removes the lowest magneto-roton branch, yielding a cleaner spectrum for $2\leq l\leq 4$, although some higher states may still be non-CFT. We remark that projection onto the quasi-hole space is mathematically equivalent to taking $U_b\to\infty$ in Eq.~\eqref{eq:hmt}. Our numerical results show that varying $U_b$ amounts to varying an irrelevant coupling and does not change the universality class. Consequently, taking $U_b\to\infty$ leaves the universal CFT physics intact, in close analogy with lowest-Landau-level projection, which underlies the fuzzy-sphere approach.

The quasi-hole-space projection greatly reduces the Hilbert space and enables us to access larger system sizes. At $N_{mf}=8$, the dimensions of the rotation-scalar, flavour-singlet spaces before and after projection are respectively 3973 and 850. However, whether or not it improves the quality of the measured conformal data remains to be tested. Since the $\SU(2)_1$-Higgs transition is not controllably accessible by other methods and is therefore difficult to benchmark, we turn to another model with better-established conformal data, namely the Ising CFT at fractional filling~\cite{Voinea2024}, for which the effectiveness of the quasi-hole-space projection can be demonstrated more clearly.

\subsection{Ising CFT Regularized via Fractional Quantum Hall States}

The original fuzzy-sphere construction of the Ising CFT is built on an integer quantum Hall background~\cite{Zhu2022}: the charge sector remains in an integer quantum Hall state, while the (pseudo-)spin sector undergoes the 3D Ising transition. Ref.~\cite{Voinea2024} later showed that the integer quantum Hall background can be replaced by a fractional quantum Hall state, either an Abelian Laughlin state or a non-Abelian Moore-Read state, since the charge sector remains gapped in these cases as well. Taking the $\nu=1/3$ Laughlin state as an example, the model involves two flavours of fermions whose number of orbitals and electrons are related by $N_m=3N_e-2$. The Hamiltonian is 
\begin{align}
    H&=V_\FQH H_\FQH+H_\Ising\nonumber\\
    H_\FQH&=\sum_{\sigma}H_{\intra,\sigma}+H_\inter\nonumber\\
    &=\sum_{\sigma}(c^\dagger_\sigma c^\dagger_\sigma)_{2s-1}\cdot(c_\sigma c_\sigma)_{2s-1}+2\sum_{l=0,1}(c^\dagger_{\downarrow} c^\dagger_{\uparrow})_{2s-l}\cdot(c_{\uparrow} c_{\downarrow})_{2s-l}\nonumber\\
    H_\Ising&=\frac{1}{2}\int\rd^2\br_1\,\rd^2\br_2 U(r_{12})\left(n^0(\br_1)n^0(\br_2)-n^x(\br_1)n^x(\br_2)\right)+h\int\rd^2\br\,n^z(\br)
\end{align}
where $H_\FQH$ is also the parent Hamiltonian of the Halperin-$(332)$ state~\cite{Halperin:1983zz} and $H_{\intra,\sigma}$ is the parent Hamiltonian of the $1/3$-Laughlin wave function on the spin-$\sigma$ layer. We use a short-hand notation
\begin{multline*}
    (c_{\sigma_1}^\dagger c_{\sigma_2}^\dagger)_{2s-l}\cdot (c_{\sigma_3} c_{\sigma_4})_{2s-l}\\=\sum_{mm_1m_2m_3m_4}c_{m_1\sigma_1}^\dagger c_{m_2\sigma_2}^\dagger c_{m_3\sigma_3}c_{m_4\sigma_4}\langle sm_1,sm_2|(2s-l)m\rangle\langle (2s-l)m|sm_3,sm_4\rangle.
\end{multline*}
Here, $U(r_{12})$ is parameterized by pseudopotentials $U_2$ and $U_3$. In Ref.~\cite{Voinea2024}, an optimal value $U_2=0.49$, $U_3=0.09$, and $h=0.135$ is reported, with $V_\FQH=1$ set as the energy unit. The charge sector remains in the $\nu=1/3$ Laughlin state across the transition, while tuning $h$ drives an Ising transition in the (pseudo-)spin sector between the $\downarrow$ flavour-polarized state and a $\BZ_2$ symmetry-broken phase.

Although the $H_\FQH$ term is not dominant in the Hamiltonian, we find that the Ising CFT states mostly reside in the zero-energy space of $H_\FQH$. We construct this space by first projecting each segment onto the quasi-hole space $\cH^\text{intra-qh}_\sigma(N_e,lm,Z)$ of the $\nu=1/3$ Laughlin wave function. Composing them gives an intralayer-projected space $\cH^{\text{intra-qh}}(N_e,lm,Z)$. After the intralayer projection, we then construct the zero-energy space $\cH^\qh$ of $H_\inter$. As $H_\inter$ conserves the number of particles separately in two flavors, we decompose $\cH^\text{intra-qh}(N_e,lm,Z)=\bigoplus_{N_{e\downarrow}N_{e\uparrow}}\cH^\text{intra-qh}(N_{e\downarrow}N_{e\uparrow},lm,Z)$ and diagonalize $H_\inter$ within each $N_{e\downarrow}N_{e\uparrow}$ sector to obtain the zero-energy space $\cH^\qh(N_{e\downarrow}N_{e\uparrow},lm,Z)$, and the final zero-energy quasi-hole space $\cH^\qh$ is their direct sum $\cH^\qh(N_e,lm,Z)=\bigoplus_{N_{e\downarrow}N_{e\uparrow}}\cH^\qh(N_{e\downarrow}N_{e\uparrow},lm,Z)$.

We measure the overlap $\langle\Phi|\Pi_\qh|\Phi\rangle$ (Table~\ref{tbl:trunc_wt}), where $\Pi_\qh$ denotes the projection onto the quasi-hole space $\cH^\qh$. For CFT states with $l\leq 2$ such as $\BI,\epsilon,\sigma,\partial^\mu\epsilon,\partial^\mu\sigma$, the weight is very close to unity, typically above 0.998 for the scalars; for spin-2 operators like $T^{\mu\nu}$, the weight is around 0.98 but increasing with system size. On the contrary, non-CFT states like the magneto-roton state (\emph{e.~g.}~the lowest spin-4 Ising-even state) have a very low weight $<0.05$ within the quasi-hole space. 

\begin{table}[htbp]
    \centering
    \renewcommand{\tabcolsep}{10pt}
    \begin{tabular}{r|cccc}
        \hline\hline
        $N_e$ & 8 & 7 & 6 & 5 \\ 
        \hline
        $\BI$ & 0.9981 & 0.9987 & 0.9989 & 0.9996 \\
        $\epsilon$ & 0.9988 & 0.9992 & 0.9993 & 0.9999 \\
        $\sigma$ & 0.9987 & 0.9991 & 0.9994 & 0.9997 \\
        $\partial^\mu\epsilon$ & 0.9930 & 0.9930 & 0.9928 & 0.9931 \\
        $\partial^\mu\sigma$ & 0.9944 & 0.9948 & 0.9945 & 0.9959 \\
        $T^{\mu\nu}$ & 0.9827 & 0.9803 & 0.9755 & 0.9711 \\
        roton & 0.0350 & 0.0107 & 0.0031 & 0.0018 \\
        \hline\hline
    \end{tabular}
    \caption{The weight of different states of Ising CFT at filling-$1/3$ within the quasi-hole space $\cH^\qh$ measured by the overlap $\langle\Phi|\Pi_\qh|\Phi\rangle$ for different system sizes. ``Roton'' denotes the lowest spin-4 Ising-even state.}
    \label{tbl:trunc_wt}
\end{table}

\begin{figure}[htbp]
    \centering
    \includegraphics[width=0.48\linewidth]{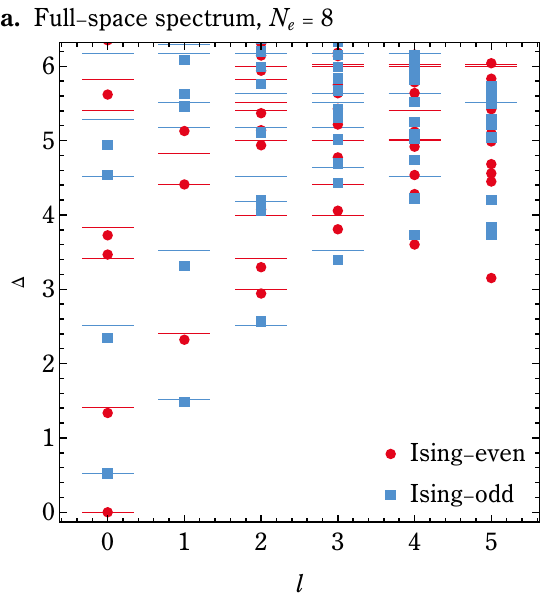}
    \includegraphics[width=0.48\linewidth]{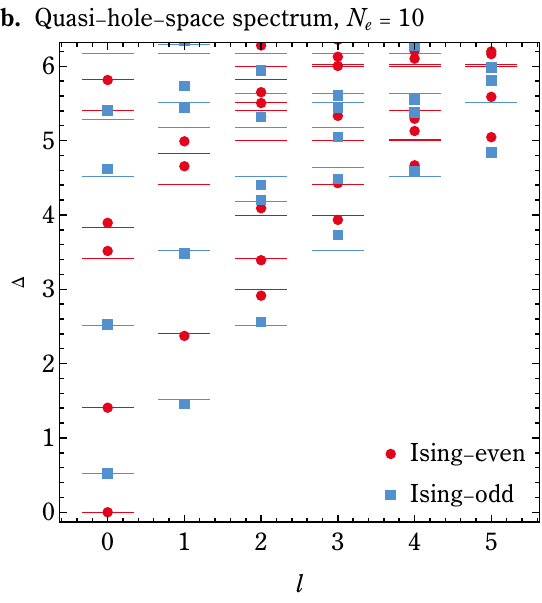}
    \includegraphics[width=0.48\linewidth]{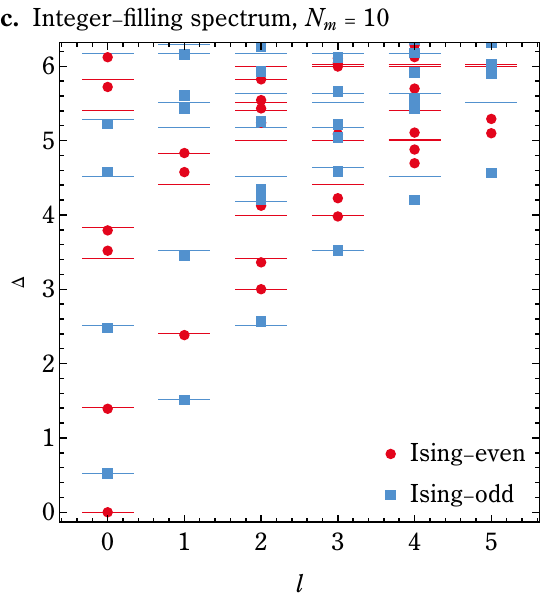}
    \caption{The spectrum in (a) the full Hilbert space for $N_e=8$ at $V_\FQH=1,U_2=0.49,U_3=0.09,h=0.135$, (b) the quasi-hole space for $N_e=10$ at filling-$1/3$ at $U_2=3.86,U_3=1,h=1.79$, in comparison with (c) the Ising-model spectrum at integer filling. The coloured bars denote the expected operator content from conformal bootstrap.}
    \label{fig:trunc_spec}
\end{figure}

We then examine the energy spectrum of $H_\Ising$ within the projected quasi-hole space $\cH^\qh$. We set $U_3=1$ and determine $U_2,h$ by optimizing a conformal cost function defined as the root mean square of the deviations of the scaling dimensions of all 10 operators with $\Delta<3.5$ from values of conformal bootstrap~\cite{Chang:2024whx}. For $N_e=10$, we find $U_2=3.86,h=1.79$. The spectrum agrees well with the expected operator content (Figure \ref{fig:trunc_spec}b) for $l\leq 4$ and $\Delta\lesssim 5.5$. The spectrum obtained from the quasi-hole space computation is of substantially higher quality than that obtained from the full Hilbert space (Figure~\ref{fig:trunc_spec}a). First, the low-lying CFT states with $l\leq 2$ and $\Delta\leq 3$ show better agreement with the bootstrap results. More importantly, within the window $l\leq 4$ and $\Delta\leq 5.5$, the full Hilbert space spectrum contains many non-CFT states, such as gapped magneto-roton excitations, whereas these non-CFT states are completely absent from the quasi-hole space spectrum. These observations demonstrate the effectiveness of the quasi-hole-space projection in this setting. As in the $\SU(2)_1$ Higgs transition, the projection becomes exact in the limit $V_\FQH\to\infty$, and it leaves the IR universality class of the transition unchanged.

We note that this projection is different from in $\SU(2)_1$-Higgs transition in that it does not start from the wave function of the FQH state in its vicinity; instead, it projects onto the quasi-hole space of the Halperin-$(332)$ state with filling $\nu_0=2/5$ that is higher than $\nu=1/3$ at the transition. The quasi-hole filling is kept a constant $\nu_\text{qh}=\nu_0\nu/2(\nu_0-\nu)=1$. This provides an alternative perspective for the Ising CFT on Laughlin states: The quasi-holes are at half filling with respect to the total filling 2 of the bilayer. The interaction drives an Ising-type transition between a quantum Hall ferromagnet state and the polarized state formed by the quasi-holes.

Remarkably, the quasi-hole space projected Ising model admits a direct mapping to the original fuzzy-sphere Ising construction based on an integer quantum Hall background. More specifically, the quasi-hole space $\cH^\qh(N_e,l,Z)$ is linearly isomorphic to the Hilbert space $\cH'(N_e,l,Z)$ of the original fuzzy-sphere Ising model with the same number of fermions $N_e$, namely a $\nu'=1$ quantum Hall bilayer with $N'_m=N_e$ orbitals. One direct piece of evidence for this is the matching of Hilbert-space dimensions: not only do the two spaces have the same total dimension, but their corresponding sectors labelled by $(N_{e\downarrow}N_{e\uparrow},l_{\downarrow}l_{\uparrow})$ also match in dimension, $\dim \cH'(N_{e\downarrow}N_{e\uparrow},l_{\downarrow}l_{\uparrow},Z)=\dim \cH^\qh(N_{e\downarrow}N_{e\uparrow},l_{\downarrow}l_{\uparrow},Z)$.

This linear isomorphism can be made explicit at the level of wave functions. Consider a state of the $\nu'=1$ quantum Hall bilayer with $N'_m=N_e$ orbitals, whose wave function takes the form $\Psi'(\{z_i\},\{w_k\})$, where $z_i$ and $w_k$ are the coordinates of the spin-down and spin-up particles on the complex plane, with $i=1,\dots,N_{\downarrow}$ and $k=1,\dots,N_{\uparrow}$. The corresponding $\nu=1/3$ state with $N_m=3N_e-2$ is obtained via flux attachment,
\begin{equation} 
    \Psi(\{z_i\},\{w_k\})=\prod_{i<j}^{N_{\downarrow}}(z_i-z_j)^2\prod_{k<l}^{N_{\uparrow}}(w_k-w_l)^2\prod_{i=1}^{N_{\downarrow}}\prod_{k=1}^{N_{\uparrow}}(z_i-w_k)^2\Psi'(\{z_i\},\{w_k\}).
\end{equation} One can show that $\Psi$ is a zero-energy state of $H_\FQH$ by rewriting it as a quasi-hole state of the Halperin-$(332)$ state, for which $H_\FQH$ is a parent Hamiltonian. Although this map defines a linear isomorphism between the two spaces, it does not preserve the inner product: two orthogonal states satisfying $\langle\Psi'_1|\Psi'_2\rangle=0$ in $\cH'$ need not be mapped to orthogonal states $\Psi_1$ and $\Psi_2$ in $\cH^\qh$.

The spectra at filling $1/3$ and filling $1$ with the same $N_e$ exhibit notable similarities (Figure~\ref{fig:trunc_spec}b and c). An interesting question is whether the particle-hole symmetry in $\cH'(N_e,l,Z)$ has a counterpart in $\cH^\qh(N_e,l,Z)$, and whether the Hamiltonian and states at filling $1/3$ admit a manifest particle-hole symmetry. A particle-hole symmetry would map the sector $(N_{e\downarrow}N_{e\uparrow},l_{\downarrow}l_{\uparrow})$ to $(N_{e\downarrow}N_{e\uparrow},l_{\uparrow}l_{\downarrow})$. As a preliminary indication of such a symmetry, we find that the ground-state wave function carries equal weight in sectors related by the exchange $l_{\downarrow}\leftrightarrow l_{\uparrow}$.

We also test the quasi-hole-space projection for Ising CFT on other FQH states. For Ising CFT on the filling-$1/m$ Laughlin state (bosonic for even $m$ and fermionic for odd-$m$), we project onto the quasi-hole space of the Halperin-$(m,m,m-1)$ state, which is the zero-energy space of 
\begin{equation*}
    H_\FQH=\sum_{\substack{\sigma,m-l\in2\BZ\\0\leq l\leq m-1}}[a^\dagger_\sigma a^\dagger_\sigma]_{2s-l}\cdot[a_\sigma a_\sigma]_{2s-l}+2\sum_{l=0,\dots,m-2}[a^\dagger_{\downarrow} a^\dagger_{\uparrow}]_{2s-l}\cdot[a_{\uparrow} a_{\downarrow}]_{2s-l}
\end{equation*}
where $a^\dagger_\sigma$ is the boson/fermion creation operator on the fuzzy sphere. This space admits a similar isomorphism with the quantum Hall bilayer at half filling
\begin{equation} 
    \Psi(\{z_i\},\{w_k\})=\prod_{i<j}^{N_{\downarrow}}(z_i-z_j)^{m-1}\prod_{k<l}^{N_{\uparrow}}(w_k-w_l)^{m-1}\prod_{i=1}^{N_{\downarrow}}\prod_{k=1}^{N_{\uparrow}}(z_i-w_k)^{m-1}\Psi'(\{z_i\},\{w_k\}).
\end{equation} 
For the filling-$1/2$ bosonic Laughlin state, we find that the CFT states $\BI,\epsilon$, and $\sigma$ have weight greater than 0.999 in the quasi-hole space (Table~\ref{tbl:trunc_wt_boson}). For the filling-1 bosonic Pfaffian state, we perform the intra-layer projection to the zero-energy space of the parent Hamiltonian of the bosonic Pfaffian state within each flavour\footnote{Further projection can be made to the quasi-hole space of a suitable non-Abelian spin-singlet FQH state~\cite{Estienne:2011}.}
\begin{equation*}
    H_\FQH=\sum_{\sigma}[b^\dagger_\sigma b^\dagger_\sigma b^\dagger_\sigma]_{3s}\cdot[b_\sigma b_\sigma b_\sigma]_{3s}.
\end{equation*}
The CFT states $\BI,\epsilon$, and $\sigma$ have weight greater than 0.96 in the quasi-hole space. Possible reasons why these weights are smaller than those for the Laughlin states include that the bosonic Pfaffian state is favoured by the density-density interaction instead of its parent Hamiltonian, and that the CFT quality is not as good as Ising CFT on Laughlin states.

\begin{table}[htbp]
    \centering
    \setlength{\tabcolsep}{10pt}
    \begin{tabular}{c|cc}
        \hline\hline
        State & Bosonic Laughlin & Bosonic Pfaffian\\[-4pt]
        & $\nu=1/2$, $N_e=8$ & $\nu=1$, $N_e=10$\\
        \hline 
        $\BI$ & 0.9992 & 0.9643 \\
        $\epsilon$ & 0.9995 & 0.9720 \\
        $\sigma$ & 0.9994 & 0.9647 \\
        \hline\hline
    \end{tabular}
    \caption{The weights of different states within the quasi-hole space $\cH^\qh$ measured by the overlap $\langle\Phi|\Pi_\qh|\Phi\rangle$ for the Ising CFT on $\nu=1/2$ bosonic Laughlin state and $\nu=1$ bosonic Pfaffian state.}
    \label{tbl:trunc_wt_boson}
\end{table}

\section{Discussion}
\label{sec:discussion}

In this work, we exploit the symmetry and Hilbert-space structures of fuzzy-sphere models to advance their use in studying 3D CFTs. Resolving the full $\SO(3)$ rotational symmetry makes larger system sizes and heavier operators accessible, while examining the many-body wave functions reveals simple organization within selected CFT states and motivates Hilbert-space projection. These two directions provide complementary ways to extract CFT data and understand how the continuum theory emerges from the microscopic model.

Our $\SO(3)$-rotation-resolving ED algorithm substantially reduces memory and time requirements, permits larger system sizes, and enables us to target specific angular-momentum sectors. It also applies to other fuzzy-sphere models, including $\SO(5)$ deconfined criticality, $\Sp(N)$ CFTs, the $\SU(2)_1$-Higgs transition, and the free Majorana fermion. Non-Abelian internal symmetries can be partially resolved as well. For the Ising CFT, using the $\SO(3)$-resolving ED, we have identified 48 spin-0 primaries up to $\Delta\approx 16.5$ including 9 parity-odd ones, 19 spin-1 primaries up to $\Delta\approx 14$ (parity-even) or $15.5$ (parity-odd), and 28 spin-2 primaries up to $\Delta\approx 12.5$. Many of these primaries were previously unknown. In several sectors, such as spin-1 parity-odd and spin-0 and spin-2 Ising-odd parity-odd, we report primaries for the first time. For the previously known operators, our results agree with conformal bootstrap, $\epsilon$-expansion, and previous fuzzy-sphere studies, and support spectral assumptions made in bootstrap.

We also identify a branch of Ising scalar primaries associated with the $\phi^n$ family, beginning with $\sigma$, $\epsilon$, $\epsilon'$, and $\sigma'$. Their wave functions have unusually large weights in a small subspace invariant under area-preserving diffeomorphisms (APD) of the sphere. These states admit a semi-classical interpretation by collective spin flip from a fully polarized configuration. For FQH transitions, we find that the low-lying CFT states lie predominantly in substantially reduced Hilbert spaces spanned by quasi-hole states. We demonstrate this first for the $\SU(2)_1$-Higgs transition and then for the Ising CFT at fractional filling, using the zero-energy spaces of the bosonic Laughlin and Halperin-$(332)$ parent Hamiltonians, respectively. Projection to these spaces reduces the computational cost and removes many non-CFT excitations while retaining the low-lying CFT spectrum.

The $\SO(3)$-symmetric algorithm could help us study problems that were previously inaccessible numerically on the fuzzy sphere. A particularly outstanding example is the finite-temperature physics of 3D CFTs~\cite{Katz:2014rla}. A CFT at finite temperature is defined on the thermal manifold $S^1_\beta\times\BR^2$. The inverse temperature $\beta$ introduces a scale and allows operators to acquire universal thermal expectation values $\langle\Phi\rangle_\beta=b_\Phi/\beta^\Delta$. It describes the experimentally accessible quantum-critical regime and provides new universal information such as free-energy densities and thermal one-point functions that is not fixed directly by the zero-temperature spectrum. 

The finite-temperature behaviour of a fuzzy-sphere model, characterized by the thermal partition function $Z(\beta)=\operatorname{tr}\exp(-\beta H)$, corresponds to a CFT on $S^1_\beta\times S^2$, which approaches $S^1_\beta\times\BR^2$ in the limit $R/\beta\to\infty$. This was previously difficult because evaluating $Z(\beta)$ requires access to high-lying excited states. The $\SO(3)$-resolution has made this task substantially easier. As evidence, at $N_m=14$ the largest sector of the Hilbert space has dimension $2\times 10^4$, making a complete diagonalization of the Hamiltonian feasible.

Access to a larger portion of the spectrum may further enable the study of real-time and non-equilibrium dynamics, in particular, the transport properties. Of particular interest is the zero-frequency conductivity at charge neutrality. In a holographic description, the ratio between dc and $\nu\to\infty$ conductivity $\sigma_{\text{dc}}/\sigma_\infty=1+4\gamma$ is determined by a coefficient $\gamma$ that parametrizes the independent tensor structure of the $JJT$ OPE~\cite{Katz:2014rla}. Directly accessing the non-equilibrium dynamics on the fuzzy sphere could provide an opportunity to verify this relation. 

ED with full rotation-symmetry resolution could also be generalized to higher dimensions. Combined with Hilbert-space truncation, this approach could enable studies of 4D and 5D CFTs on the fuzzy four-sphere and its equator~\cite{Zhao2025,Meng2026,Xue2026,Qian2026Sep} at larger system sizes. These developments could help address outstanding questions, including the extent of the conformal window in 4D gauge theories~\cite{Banks:1981nn} and the existence of interacting non-Lagrangian fixed points above the upper critical dimension.

It is also interesting to further investigate the organizing principles of CFT states within the fuzzy-sphere Hilbert space. One direction is to better understand the $\phi^n$ branch from the perspective of APD, as well as its connection to semi-classical descriptions in QFT~\cite{Sannino:2026ena}. For the quasi-hole-space projection, we expect this to be a general strategy for studying CFTs intertwined with fractional quantum Hall phases. An important question is whether the projected space admits additional exact symmetries that are absent in the unprojected space, such as particle-hole symmetry. More broadly, it would be interesting to identify the natural algebraic structure within the quasi-hole space, which may differ substantially from the fuzzy-sphere algebra of the full Hilbert space~\cite{He2025Jun,Eck2026}. Given the close relation between FQH quasi-hole spaces and quantum Hall edge theories~\citep{Read:2007cv}, this structure may also be connected to the corresponding edge algebras and could provide new clues to the algebraic organization of 3D CFT states. Finally, it would be worthwhile to explore whether there exist further physically motivated Hilbert-space projections or truncations that could facilitate both the numerical and theoretical study of CFTs on the fuzzy sphere.

\section*{Acknowledgements}

We would like to thank Alexander Frenkel, Davide Gaiotto, Johan Henriksson, Emmanuel Katz, Justin Kulp, Neville Rajappa, Chong Wang, and Wei Zhu for fruitful discussions. Z.~Z.~acknowledges support from the Natural Sciences and Engineering Research Council of Canada (NSERC) through Discovery Grants. Research at Perimeter Institute is supported in part by the Government of Canada through the Department of Innovation, Science and Industry Canada and by the Province of Ontario through the Ministry of Colleges and Universities. Y.~C.~H.~thanks IHES
for its hospitality during the completion of this work. Although artificial intelligence tools, including ChatGPT 6.0 Astra and Claude Fable 5.0, provided assistance during exploration and writing, the authors alone developed the code and obtained the results presented in the main text. The code used to generate the results in this paper is publicly available in the package \texttt{FuzzifiED}~\cite{FuzzifiED}.

\paragraph{Note added} While preparing this manuscript, we became aware of a parallel study~\cite{Baran:2026opo} that presents a similar $\SO(3)$-resolving ED algorithm.

\appendix

\section{Detail for Conformal Generators of Ising CFT}

We assume the conformal generator has the same form as the first moment of the Hamiltonian density,
\begin{align}
    \Lambda&=U'_0\Lambda_{U,0}+U'_1\Lambda_{U,1}+h'(n^z)_1-\mu'(n^0)_1\\
    \Lambda_{U,0}&=[(c^\dagger_{\downarrow}c^\dagger_{\uparrow})_{2s}(c_{\uparrow}c_{\downarrow})_{2s}]_1\nonumber\\
    \Lambda_{U,1}&=\sum_{\sigma\sigma'}[(c^\dagger_\sigma c^\dagger_\sigma)_{2s-1}(c_{\sigma'} c_{\sigma'})_{2s-1}]_1\nonumber
\end{align}
where $(\Phi)_l$ denotes taking the angular-momentum-$l$ component of $\Phi$. The extra term $(n^0)_1$ is proportional to the angular momentum $L$; its $l=0$ component $(n^0)_0$ reduces to the conserved electric charge. 

We determine the coefficients $U'_0$, $U'_1$, and $h'$ by minimizing $\Lambda\|\BI\rangle$ while keeping $\langle\partial\epsilon\|\Lambda\|\epsilon\rangle=\sqrt{2\Delta_\epsilon}$. As $(n^0)_1$ annihilates the scalars, this scheme leaves $\mu'$ ambiguous. We determine $\mu'$ by requiring that $\langle\Phi_l\|K\|\Phi_l\rangle=0$ for an arbitrary spinning operator $\Phi_l$.

\begin{figure}[htbp]
    \centering
    \includegraphics[width=0.32\linewidth]{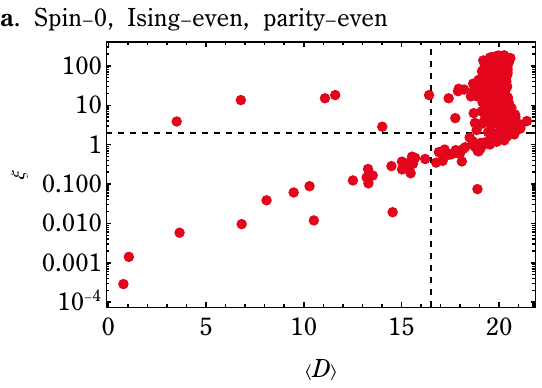}
    \includegraphics[width=0.32\linewidth]{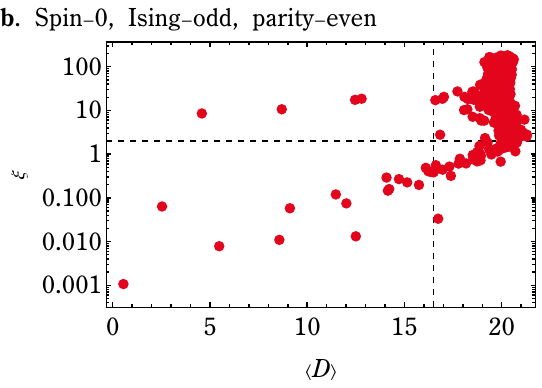}
    \includegraphics[width=0.32\linewidth]{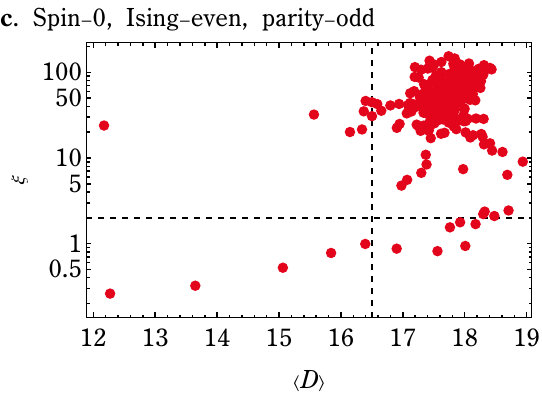}
    \includegraphics[width=0.32\linewidth]{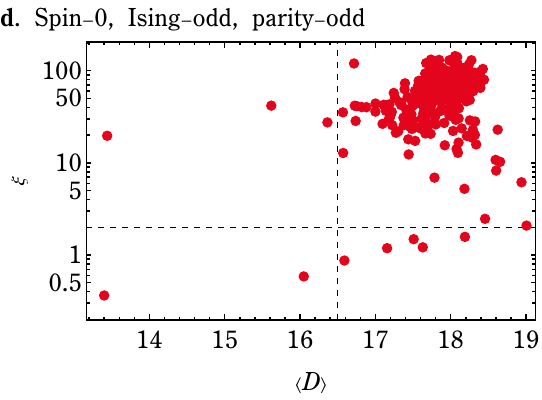}
    \includegraphics[width=0.32\linewidth]{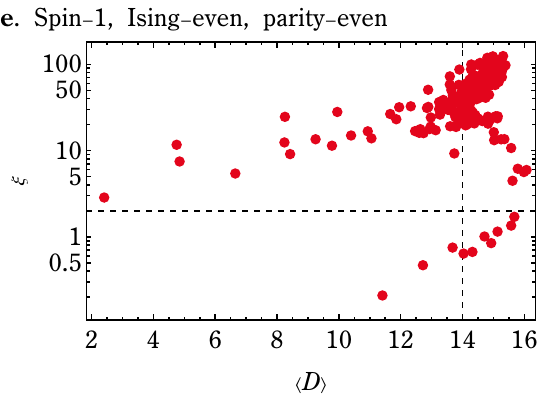}
    \includegraphics[width=0.32\linewidth]{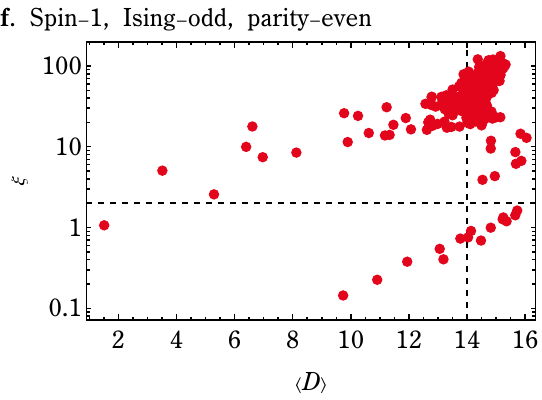}
    \includegraphics[width=0.32\linewidth]{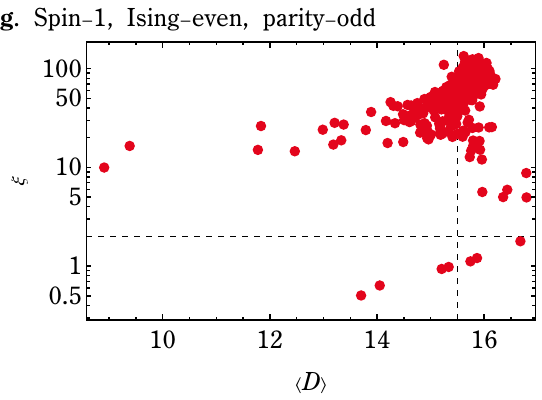}
    \includegraphics[width=0.32\linewidth]{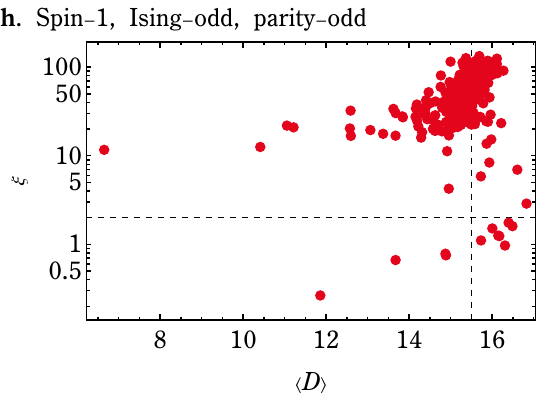}
    \includegraphics[width=0.32\linewidth]{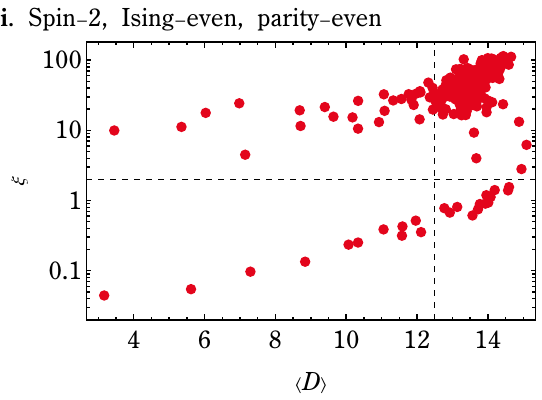}
    \includegraphics[width=0.32\linewidth]{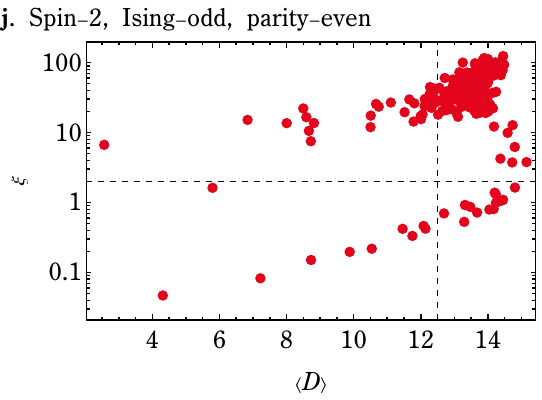}
    \includegraphics[width=0.32\linewidth]{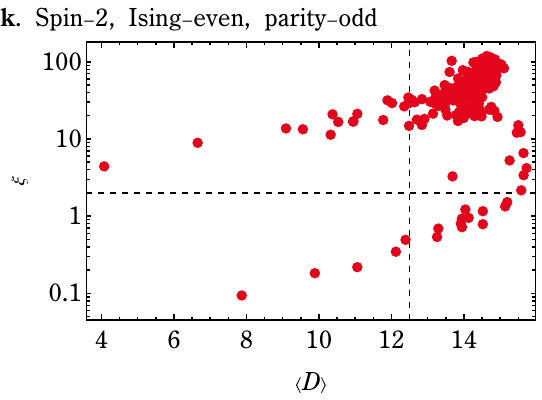}
    \includegraphics[width=0.32\linewidth]{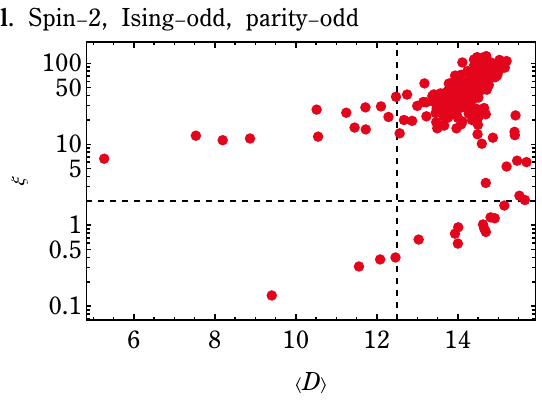}
    \caption{Eigenvalues $\xi_\alpha$ of $\Xi=|K|^2$ in the various sectors, plotted against the expectation value of the dilatation operator $\langle D\rangle=\langle\xi_\alpha|D|\xi_\alpha\rangle$. The grid lines denote the cut-off values for $\xi$ and scaling dimension $\Delta$.}
    \label{fig:k2eig}
\end{figure}

\begin{figure}[htbp]
    \centering
    \includegraphics[width=0.32\linewidth]{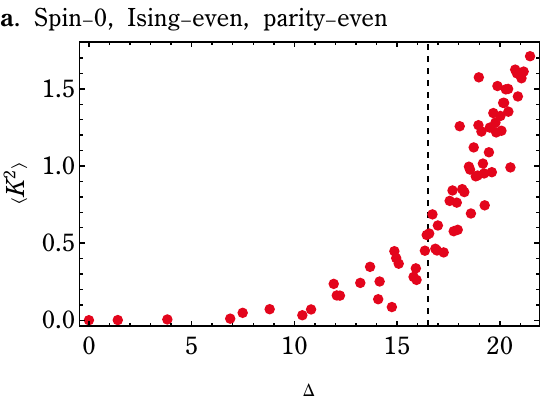}
    \includegraphics[width=0.32\linewidth]{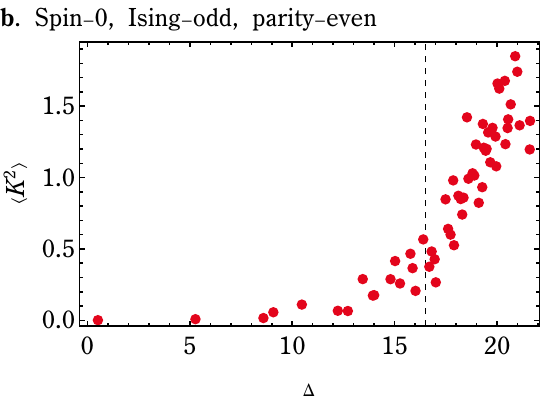}
    \includegraphics[width=0.32\linewidth]{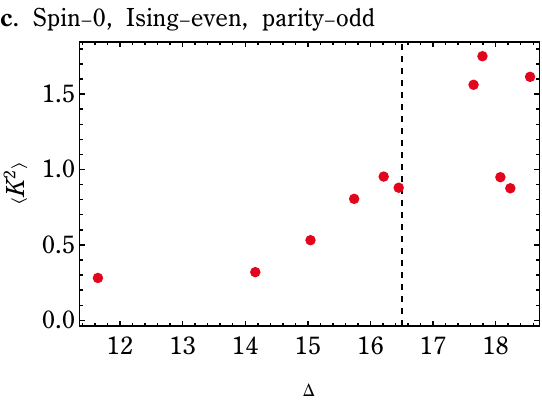}
    \includegraphics[width=0.32\linewidth]{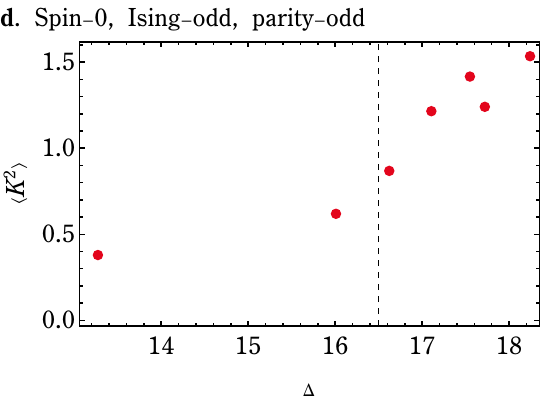}
    \includegraphics[width=0.32\linewidth]{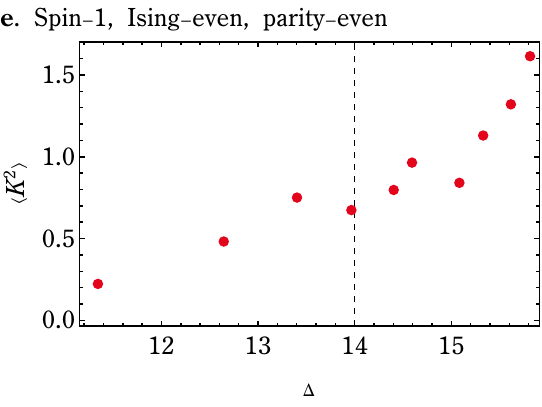}
    \includegraphics[width=0.32\linewidth]{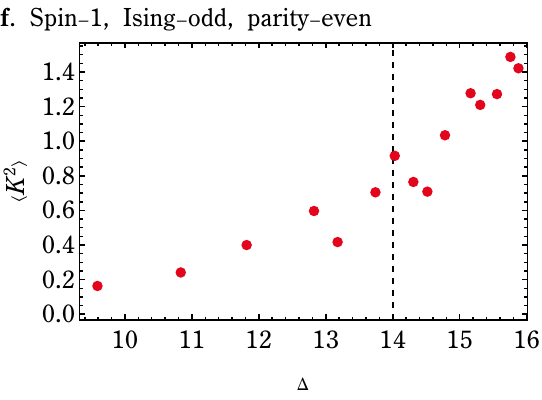}
    \includegraphics[width=0.32\linewidth]{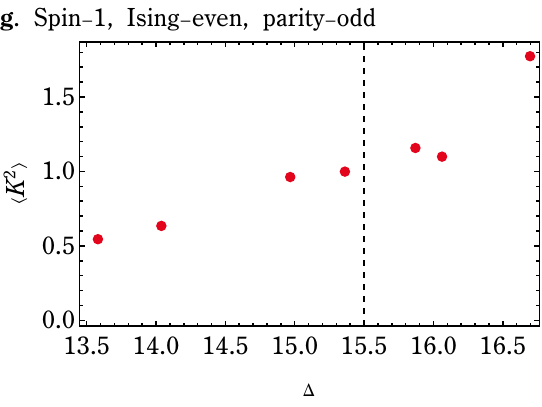}
    \includegraphics[width=0.32\linewidth]{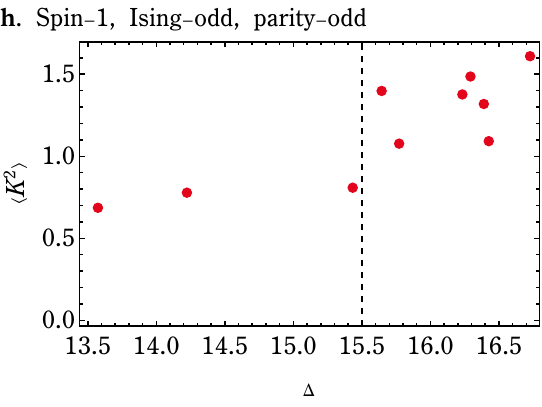}
    \includegraphics[width=0.32\linewidth]{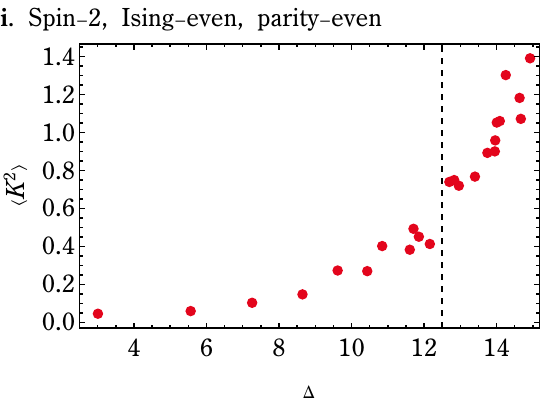}
    \includegraphics[width=0.32\linewidth]{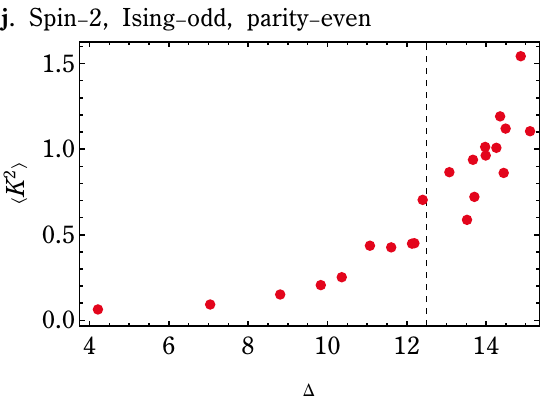}
    \includegraphics[width=0.32\linewidth]{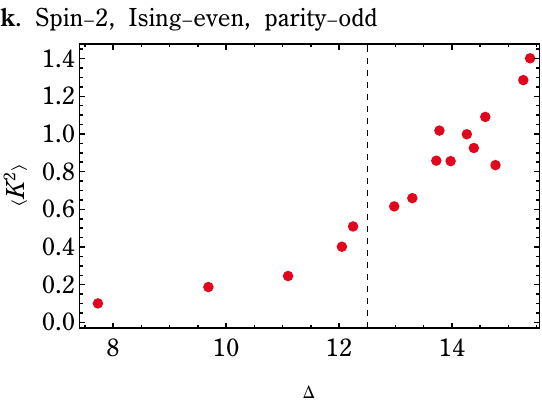}
    \includegraphics[width=0.32\linewidth]{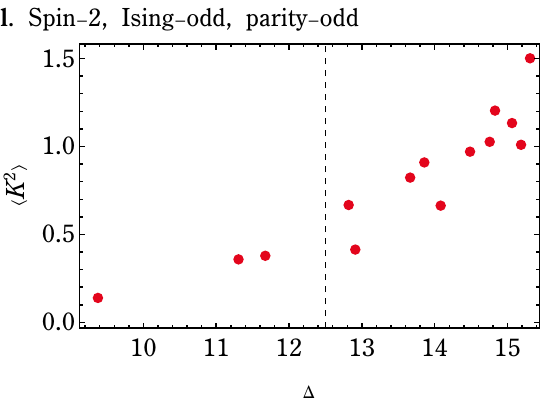}
    \caption{Scaling dimensions $\Delta$ and expectation values $\langle \Xi\rangle$ of the identified primaries in the various sectors.}
    \label{fig:deig}
\end{figure}

To calculate the expectation value $\langle\Phi_l\|\Xi\|\Phi_l\rangle$ of $\Xi=|K|^2=P_\mu K^\mu$ on a spin-$l$ state, we first calculate the action of $K$ on $\|\Phi_l\rangle$ decomposed into different spin channels
\begin{align*}    
    \|(K\Phi)_{l'}\rangle&=\left(K\|\Phi_l\rangle\right)_{l'}&(l'&=l,l\pm1)
\end{align*}
We then evaluate the inner product by choosing a representative component $|\Phi_{l0}\rangle$.
\begin{align}
    \langle\Phi_l\|\Xi\|\Phi_l\rangle&=\sum_{m=0,\pm1}\langle\Phi_{l,0}|(K_m)^\dagger K_m|\Phi_{l,0}\rangle\nonumber\\
    &=    \sum_{\substack{l'=l,l\pm1\\m=0,\pm1}}\langle\Phi_{l,0}|(K_m)^\dagger\left[\operatorname{tr}_{\Phi'_{l'm}}|\Phi'_{l',m}\rangle\langle\Phi'_{l',m}|\right]K_m|\Phi_{l,0}\rangle\nonumber\\
    &=\sum_{\substack{l'=l,l\pm1\\m=0,\pm1}}\operatorname{tr}_{\Phi'_{l'm}}|\langle\Phi'_{l',m}|K_m|\Phi_{l,0}\rangle|^2\nonumber\\
    &=\sum_{\substack{l'=l,l\pm1\\m=0,\pm1}}\operatorname{tr}_{\Phi'_{l'}}|\langle l'm|1m,l0\rangle\langle\Phi'_{l'}|K|\Phi_{l}\rangle|^2\nonumber\\
    &=\sum_{\substack{l'=l,l\pm1}}\left[ \sum_{m=0,\pm1}|\langle l'm|1m,l0\rangle|^2\right]\left\|\|(K\Phi)_{l'}\rangle\right\|^2\nonumber\\
    &=\left\|(K\Phi)_{l-1}\rangle\right\|^2\left[\left(\sqrt{\frac{l-1}{2(2l+1)}}\right)^2+\left(-\sqrt{\frac{l}{2l+1}}\right)^2+\left(\sqrt{\frac{l-1}{2(2l+1)}}\right)^2\right]\nonumber\\
    &\qquad\qquad+\left\|(K\Phi)_{l}\rangle\right\|^2\left[\left(\frac{1}{\sqrt{2}}\right)^2+0^2+\left(-\frac{1}{\sqrt{2}}\right)^2\right]\nonumber\\
    &\qquad\qquad+\left\|(K\Phi)_{l+1}\rangle\right\|^2\left[\left(\sqrt{\frac{l+2}{2(2l+1)}}\right)^2+\left(\sqrt{\frac{l+1}{2l+1}}\right)^2+\left(\sqrt{\frac{l+2}{2(2l+1)}}\right)^2\right]\nonumber\\
    &=\frac{2l-1}{2l+1}\left\|(K\Phi)_{l-1}\rangle\right\|^2+\left\|(K\Phi)_{l}\rangle\right\|^2+\frac{2l+3}{2l+1}\left\|(K\Phi)_{l+1}\rangle\right\|^2\nonumber\\
    &=\sum_{l'=l,l\pm1}\frac{2l'+1}{2l+1}\left\|(K\Phi)_{l'}\rangle\right\|^2
\end{align}
For Lorentz scalars with $l=0$, only the $l'=1$ survives 
\begin{equation*}
    \langle\Phi_0\|\Xi\|\Phi_0\rangle=3\left\|\|(K\Phi)_1\rangle\right\|^2
\end{equation*}

We plot the eigenvalues of $\Xi=|K|^2$ in the various sectors against the expectation value of the dilatation operator in Figure~\ref{fig:k2eig} and scaling dimensions $\Delta$ and expectation values $\langle \Xi\rangle$ of the identified primaries in Figure~\ref{fig:deig}.

\section{More Discussions on Area-Preserving Diffeomorphism}

The fuzzy sphere Ising model is built with fermions $c_{m\sigma}$, where $m = -s, -s+1, \cdots, s$ is in the spin-$s$ representation of the $\SO(3)_r$ sphere rotation, and $\sigma = \uparrow,\downarrow$ is the flavour degree of freedom providing the Ising $\BZ_2$ symmetry. We consider the case with $N_m = 2s+1$ fermions, so the many-body Hilbert space is
\begin{equation}\label{eq:HS}
    \prod_{i=1}^{N_m} c^\dag_{m_i,\sigma_i} |0\rangle .
\end{equation}
Although the fuzzy sphere model only has $\SO(3)_r$ sphere rotation and Ising $\BZ_2$ symmetry, it is helpful to consider an enlarged symmetry $\SU(N_m)\times \SU(2)_f$, where the fermions are in the bi-fundamental of $\SU(N_m)$ and $\SU(2)_f$. The $\SU(N_m)$, at $N_m\rightarrow\infty$, corresponds to the area-preserving diffeomorphism (APD) of the (fuzzy) sphere~\cite{Hoppe:1986aj,Swain:2004hp}. The states invariant under APD are $\SU(N_m)$ singlets, and there are in total $N_m+1$ such states, which form a spin-$N_m/2$ multiplet of $\SU(2)_f$. These states have trivial correlation functions because the constant function is the only function on $S^2$ that is APD-invariant. 

It is convenient to decompose the Hilbert space \eqref{eq:HS} into the irreps of $\SU(N_m)\times \SU(2)_f$. Due to the antisymmetry of fermions, the non-vanishing fermion states have dual irreps in $\SU(N_m)$ and $\SU(2)_f$. Specifically, we can write the spin-$j$ irrep of $\SU(2)_f$ using the Young diagram $(N_m/2 + j,\, N_m/2 - j)$
\begin{equation*}
    \begin{tikzpicture}[x=0.4cm,y=0.4cm]
        \foreach \x in {0,1,5,6}
            \draw (\x,1) rectangle ++(1,1);
        \node at (3,1.5) {$\cdots$};
        \foreach \x in {0,1,4}
            \draw (\x,0) rectangle ++(1,1);
        \node at (3,0.5) {$\cdots$};
        \draw[decorate,decoration={brace,amplitude=4pt}]
            (0,2.2) -- (7,2.2)
            node[midway,above=5pt] {$N_m/2+j$};
        \draw[decorate,decoration={brace,mirror,amplitude=4pt}]
            (0,-0.2) -- (5,-0.2)
            node[midway,below=5pt] {$N_m/2-j$};
    \end{tikzpicture}
\end{equation*}
Its dual irrep in $\SU(N_m)$ is $(2^{N_m/2 - j},\, 1^{2j})$
\begin{equation*}
    \begin{tikzpicture}[x=0.4cm,y=0.4cm]
        \foreach \y in {0,-1,-5,-6}
        \draw (0,\y) rectangle ++(1,-1);
        \node at (0.5,-3.5) {$\vdots$};
        \foreach \y in {0,-1,-4}
        \draw (1,\y) rectangle ++(1,-1);
        \node at (1.5,-3) {$\vdots$};
        \draw[decorate,decoration={brace,amplitude=4pt}]
        (-0.2,-7) -- (-0.2,0)
        node[midway,left=5pt] {$\dfrac{N_m}{2}+j$};
        \draw[decorate,decoration={brace,amplitude=4pt}]
        (2.2,0) -- (2.2,-5)
        node[midway,right=5pt] {$\dfrac{N_m}{2}-j$};
    \end{tikzpicture}
\end{equation*}
The entire Hilbert space can be written as,
\begin{equation}
    \bigoplus_{j=0}^{N_m/2}  \left(2^{\frac {N_m} 2 - j},\, 1^{2j}\right)_{\SU(N_m)} \otimes \left(\frac {N_m} 2 + j,\, \frac {N_m} 2 - j\right)_{\SU(2)_f}.
\end{equation}
We note that for each $j$ there is no multiplicity. $j = N_m/2$ corresponds to the $\SU(N_m)$ singlet and $\SU(2)_f$ spin $N_m/2$, and these states are APD-invariant. $j=N_m/2-1$ corresponds to the $\SU(N_m)$ adjoint and $\SU(2)_f$ spin-$N_m/2-1$. It is natural to associate $N_m/2 - j$ with the strength of APD violation.

\begin{table}[htbp]
    \centering
    \setlength{\tabcolsep}{10pt}
    \begin{tabular}{cc|cc|cc}
        \hline\hline
        $j$ & Number of scalars & $j$ & Number of scalars & $j$ & Number of scalars \\
        \hline 
        $26$ & $1$                   & $17$ & $1.23 \times 10^{13}$ & $ 8$ & $6.46 \times 10^{20}$ \\
        $25$ & $0$                   & $16$ & $1.99 \times 10^{14}$ & $ 7$ & $1.86 \times 10^{21}$ \\
        $24$ & $25$                  & $15$ & $2.58 \times 10^{15}$ & $ 6$ & $4.52 \times 10^{21}$ \\
        $23$ & $1827$                & $14$ & $2.70 \times 10^{16}$ & $ 5$ & $9.20 \times 10^{21}$ \\
        $22$ & $1.90 \times 10^{5}$  & $13$ & $2.30 \times 10^{17}$ & $ 4$ & $1.56 \times 10^{22}$ \\
        $21$ & $1.30 \times 10^{7}$  & $12$ & $1.62 \times 10^{18}$ & $ 3$ & $2.16 \times 10^{22}$ \\
        $20$ & $6.31 \times 10^{8}$  & $11$ & $9.46 \times 10^{18}$ & $ 2$ & $2.39 \times 10^{22}$ \\
        $19$ & $2.24 \times 10^{10}$ & $10$ & $4.61 \times 10^{19}$ & $ 1$ & $1.91 \times 10^{22}$ \\
        $18$ & $5.98 \times 10^{11}$ & $ 9$ & $1.88 \times 10^{20}$ & $ 0$ & $7.35 \times 10^{21}$ \\
        \hline\hline 
    \end{tabular}
    \caption{The number of $\SO(3)$-rotation scalars that carry the $\SU(N_m)$ representation labelled by $j$ for $N_m=52$. Note that each scalar in the table is also a spin-$j$ multiplet of $\SU(2)_f$, and the multiplicity factor $2j+1$ is not included in the count.}
    \label{tbl:sun_dim}
\end{table}

\begin{figure}
    \centering
    \includegraphics[width=0.99\linewidth]{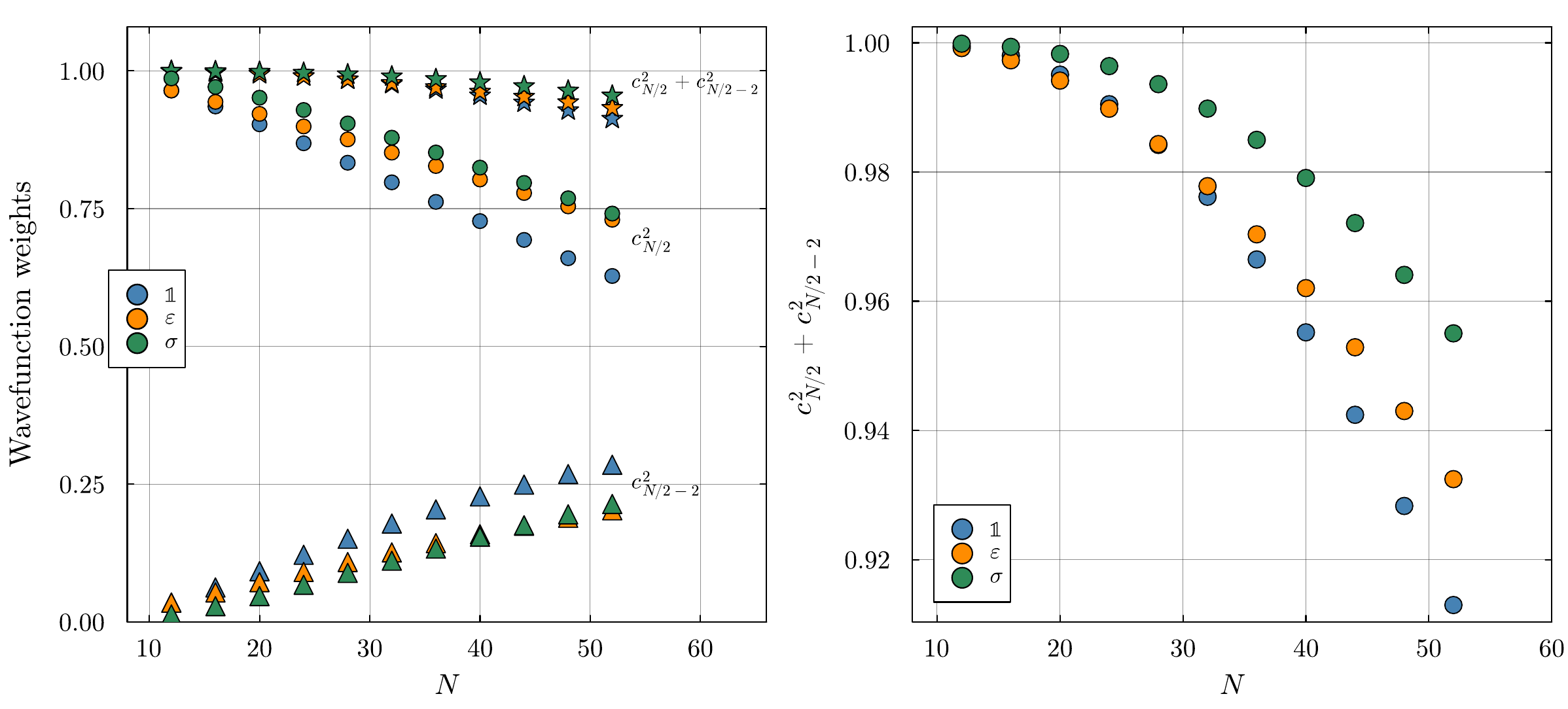}
    \caption{Wave function weights of the ground state, $\sigma$ and $\epsilon$ in the $j=N_m/2, N_m/2-2$ sectors.}
    \label{fig:WF_weights}
\end{figure}

Eventually, we need to branch $\SU(N_m)$ to the sphere $\SO(3)_r$ symmetry, and let us focus on the $\SO(3)_r$ scalar. The $\SU(N_m)$ singlet branches to an $\SO(3)_r$ scalar, while the $\SU(N_m)$ adjoint ($j = N_m/2 - 1$) yields 0 $\SO(3)_r$ scalars after branching, and $j=N_m/2-2$ yields $N_m/2-1$ $\SO(3)_r$ scalars. The higher irreps follow more complicated branching rules, which can be computed case by case for a given $N_m$ and $j$. Table~\ref{tbl:sun_dim} gives an example of $N_m=52$. There are a total of $6.79\times10^{23}$ $\SO(3)_r$ scalars, which obviously grow exponentially with $N_m$. The high-$j$ states constitute a negligible portion of all the $\SO(3)_r$ scalars. Specifically, the sectors with $j = N_m/2$ and $j = N_m/2 - 2$ contain a total of $N_m + 1 + (N_m - 3)(N_m/2 - 1) = N_m(N_m-3)/2 + 4$ $\SO(3)_r$ scalars. Nevertheless, these high-$j$ states dominate the wave functions of the CFT ground state and the $\phi^n$ states. To have a quantitative description, we can decompose the wave function of ground state, $\sigma$ and $\epsilon$ into each $j$-sector,
\begin{equation}
    |\psi\rangle = \sum_{j=0}^{N_m/2} c_j |j\rangle,
\end{equation}
the wave function is normalized as $\sum c_j^2=1$. Figure~\ref{fig:WF_weights} shows $c_{N_m/2}^2$ and $c_{N_m/2-2}^2$. Especially, one can notice that $c_{N_m/2}^2+c_{N_m/2-2}^2$ is larger than 0.9 for all the system sizes we have computed $N_m\le 52$ using DMRG. This is quite surprising, given that for $N_m=52$ there are only 1278 $\SO(3)$ scalars in the $j=N_m/2$ and $N_m/2-2$ space, which is negligible compared to the total number of scalars, $6.79\times 10^{23}$. 

It would be interesting to explore the role of APD in fuzzy sphere CFTs. One possible clue could be provided by the properties of the APD generators, $n_{lm}^0=\int_{S^2} \rd\br\, n^0(\br)Y_{lm}(\br)$. The $l=0$ and $l=1$ components are conserved quantities of the fuzzy sphere model: the former is the electric charge, which has no counterpart in the CFT, while the latter are the $\SO(3)$ angular-momentum generators of the CFT. No single CFT operator can simultaneously account for both sets of conserved quantities.\footnote{We thank Emmanuel Katz for discussions on this point.} We therefore expect $n^0(\br)$ to decouple from the IR CFT, as also supported by our numerics. In particular, its connected correlation functions with CFT operators should decay exponentially in the IR,
\begin{equation}
\left\langle n^0(r_0) O_1(r_1)\cdots O_k(r_k)\right\rangle_c \sim \exp\left(-\frac{\min_{1\leq j\leq k}|r_0-r_j|}{\xi}\right).
\end{equation}
An interesting question is whether the decoupling of $n^0(\br)$ imposes constraints on its higher-$l$ components $n^0_{lm}$ with $l\geq2$, analogous to those associated with the conserved $l=0$ and $l=1$ components.

\providecommand{\href}[2]{#2}\begingroup\raggedright\endgroup

\end{document}